\documentclass[twocolumn]{revtex4-1}
\usepackage{amsfonts,amssymb,amsmath} 
\usepackage{adjustbox}
\usepackage{mathtools}
\usepackage{graphicx}
\usepackage{xcolor}
\usepackage{bbm}
\usepackage[caption=false]{subfig}  
\usepackage{faktor}
\usepackage{xfrac}
\usepackage{tikz}
\usetikzlibrary{quantikz}

\newtheorem{lemma}{Lemma}
\newtheorem{theorem}{Theorem}
\newtheorem{corollary}{Corollary}
\newtheorem{proposition}{Proposition}
\newenvironment{proof}[1][Proof]{\par\noindent{\itshape #1.}\ }{\hfill$\square$\par\medskip}
\usepackage{color}
\definecolor{Gray}{gray}{0.9}

\definecolor{zero_Set1}{rgb}{0.894118,0.101961,0.109804}
\definecolor{one_Set1}{rgb}{0.215686,0.494118,0.721569}
\definecolor{two_Set1}{rgb}{0.301961,0.686275,0.290196}
\definecolor{three_Set1}{rgb}{0.596078,0.305882,0.639216}
\definecolor{four_Set1}{rgb}{1.000000,0.498039,0.000000}
\definecolor{five_Set1}{rgb}{1.000000,1.000000,0.200000}
\definecolor{six_Set1}{rgb}{0.650980,0.337255,0.156863}
\definecolor{seven_Set1}{rgb}{0.968627,0.505882,0.749020}
\definecolor{eight_Set1}{rgb}{0.600000,0.600000,0.600000}

\newcommand{\error}[2]{\text{\color{#2}{$#1$}}}
\newcommand{\be}{\begin{equation}}
\newcommand{\ee}{\end{equation}}
\newcommand{\bea}{\begin{eqnarray}}
\newcommand{\eea}{\end{eqnarray}}

\newcommand{\benonum}{\begin{equation*}}
\newcommand{\eenonum}{\end{equation*}}
\newcommand{\tp}{\intercal}

\newcommand{\numL}{L}  
\newcommand{\numQ}{n} 
\newcommand{\qk}{k} 
\newcommand{\lk}{M}

\newcommand{\numStab}{\numQ - \qk} 

\newcommand{\nn}{N}
\newcommand{\card}[1]{ | \, #1 \, |}

\newcommand{\idxset}[1]{\left\{1, 2, \, \dots \, #1\right\}}

\newcommand{\idxS}{i}

\newcommand{\idxR}{r}

\newcommand{\lred}{R} 

\newcommand{\lparams}{ \left[  \numL,\lk,\ld \right] }

\newcommand{\QEC}[3]{\left[ \left[  #1, #2, #3 \right]\right] }
\newcommand{\colorcode}{\left[\left[  17,1,5 \right]\right]}
\newcommand{\steanecode}{\left[\left[ 7,1,3 \right]\right]}

\newcommand{\golaycode}{\left[\left[  23,1,7 \right]\right]}

\newcommand{\ld}{d_{\lcode}}
\newcommand{\qd}{d_{\qcode}}
\newcommand{\ltt}{t_{\lcode}}
\newcommand{\qtt}{t_{\qcode}}

\newcommand{\definedby}{\coloneqq}

\newcommand{\epArg}[1]{\mathbf #1} 
\newcommand{\ep}{\mathbf E} 
\newcommand{\epE}{\mathbf E} 
\newcommand{\epX}{\mathbf X} 
\newcommand{\epZ}{\mathbf Z} 
\newcommand{\epN}{\mathbf N} 
\newcommand{\epM}{\mathbf M} 
\newcommand{\epT}{\mathbf T} 
\newcommand{\epi}[1]{\mathrm{\bf e}_{#1}}
\newcommand{\wt}[1]{wt(#1)} 
\newcommand{\colwt}[1]{wt_c(#1)} 

\newcommand{\errclass}{\mathbb E}

\newcommand{\Pauli}{{\mathcal P}} 
\newcommand{\stabgroup}{{\mathcal S}} 
\newcommand{\stab}{s} 
\newcommand{\stabblock}{{\cal G}}
\newcommand{\stabH}{{\mathcal H}}
\newcommand{\id}{\mathbf I} 
\newcommand{\normH}{{\cal N}\hspace{-2pt}\left(\stabH\right)}
\newcommand{\normS}{{\cal N}\hspace{-3pt}\left(\stabgroup\right)}
\newcommand{\qcode}{{\mathcal Q}}
\newcommand{\lcode}{{\mathcal C}}
\newcommand{\kron}{{\otimes}}

\newcommand{\syndCQ}{\Xi}
\newcommand{\syndQ}{\Sigma}
\newcommand{\syndL}{\Theta}

\newcommand{\hqx}{H^X_\qcode}
\newcommand{\hqz}{H^Z_\qcode}
\newcommand{\hq}{{H}_\qcode}
\newcommand{\hl}{{H}_\lcode}
\renewcommand{\gg}{{\mathbf G}}

\renewcommand{\pl}{{\mathbf A}}

\newcommand{\Xu}{X(u)}
\newcommand{\Xbeta}{X(\beta)}
\newcommand{\Zv}{Z(v)}
\newcommand{\Zalpha}{Z(\alpha)}

\newcommand{\Hthreequbitmat}{
\left[
\begin{array}{ccc}
1&1&0\\
1&0&1\\
\end{array}
\right]
}

\newcommand{\Hhamming}{
\left[
\begin{array}{ccccccc}
1&0&0&1&1&1&0\\
0&1&0&0&1&1&1\\
0&0&1&1&1&0&1
\end{array}
\right]
}

\newcommand{\PkronZZI}{
\left[
\begin{array}{c c c c c c c c c c c c}
1&1&0&1&1&0&1&1&0&0&0&0\\
0&0&0&1&1&0&1&1&0&1&1&0\\
1&1&0&1&1&0&0&0&0&1&1&0\\
\end{array}
\right]
}

\newcommand{\PkronZIZ}{
\left[
\begin{array}{c c c c c c c c c c c c}
1&0&1&1&0&1&1&0&1&0&0&0\\
0&0&0&1&0&1&1&0&1&1&0&1\\
1&0&1&1&0&1&0&0&0&1&0&1\\
\end{array}
\right]
}

\newcommand{\syndthree}{\begin{tabular}{|c| c c c c|}
\hline
101000 & $\color{zero_Set1}{X_{2}}$   &   $X_{8}X_{11}$   &   $X_{1}X_{9}X_{12}$   &   $X_{3}X_{7}X_{10}$
\\ 
000101 & $X_{3}$   &   $X_{9}X_{12}$   &   $X_{1}X_{8}X_{11}$   &   $X_{2}X_{7}X_{10}$
\\ 
110000 & $X_{8}$   &   $X_{2}X_{11}$   &   $X_{1}X_{9}X_{10}$   &   $X_{3}X_{7}X_{12}$
\\ 
000110 & $X_{9}$   &   $X_{3}X_{12}$   &   $X_{1}X_{8}X_{10}$   &   $X_{2}X_{7}X_{11}$
\\ 
011000 & $X_{11}$   &   $X_{2}X_{8}$   &   $X_{1}X_{7}X_{12}$   &   $X_{3}X_{9}X_{10}$
\\ 
000011 & $X_{12}$   &   $X_{3}X_{9}$   &   $X_{1}X_{7}X_{11}$   &   $X_{2}X_{8}X_{10}$
\\ 
111000 & $X_{5}$   &   $X_{1}X_{5}X_{7}X_{10}$   &   $X_{2}X_{5}X_{8}X_{11}$   &   $X_{3}X_{5}X_{9}X_{12}$
\\ 
000111 & $X_{6}$   &   $X_{1}X_{6}X_{7}X_{10}$   &   $X_{2}X_{6}X_{8}X_{11}$   &   $X_{3}X_{6}X_{9}X_{12}$
\\ 
101101 & $X_{1}$   &   $X_{7}X_{10}$   &   $X_{2}X_{9}X_{12}$   &   $X_{3}X_{8}X_{11}$
\\ 
110110 &  $\color{one_Set1}{X_{7}}$    &    $\color{two_Set1}{X_{1}X_{10}}$   &   $X_{2}X_{9}X_{11}$   &   $X_{3}X_{8}X_{12}$
\\ 
011011 & $X_{10}$   &   $X_{1}X_{7}$   &   $X_{2}X_{8}X_{12}$   &   $X_{3}X_{9}X_{11}$
\\ 
111111 & $X_{4}$   &   $X_{1}X_{4}X_{7}X_{10}$   &   $X_{2}X_{4}X_{8}X_{11}$   &   $X_{3}X_{4}X_{9}X_{12}$
\\ 
\hline 
\end{tabular} 
}

\newcommand{\ZZIsorted}{\begin{tabular}{|c| c c c c c c c c c c|}
\hline
000 & $ $ & $X_{3}$ & $X_{6}$ & $X_{9}$ & $X_{12}$ & $X_{1}X_{2}$ & $X_{4}X_{5}$ & $X_{7}X_{8}$ & $X_{1}X_{7}X_{10}$ & $X_{1}X_{7}X_{11}$
\\ 
101 & $X_{1}$ & $X_{1}X_{3}$ & $X_{1}X_{6}$ & $X_{1}X_{9}$ & $X_{1}X_{12}$ & $X_{2}$ & $X_{1}X_{4}X_{5}$ & $X_{1}X_{7}X_{8}$ & $X_{7}X_{10}$ & $X_{7}X_{11}$
\\ 
110 & $X_{7}$ & $X_{3}X_{7}$ & $X_{6}X_{7}$ & $X_{7}X_{9}$ & $X_{7}X_{12}$ & $X_{1}X_{2}X_{7}$ & $X_{4}X_{5}X_{7}$ & $X_{8}$ & $X_{1}X_{10}$ & $X_{1}X_{11}$
\\ 
011 & $X_{10}$ & $X_{11}$ & $X_{6}X_{10}$ & $X_{9}X_{10}$ & $X_{10}X_{12}$ & $X_{1}X_{2}X_{10}$ & $X_{4}X_{5}X_{10}$ & $X_{7}X_{8}X_{10}$ & $X_{1}X_{7}$ & $X_{1}X_{7}X_{10}X_{11}$
\\ 
111 & $X_{4}$ & $X_{3}X_{4}$ & $X_{4}X_{6}$ & $X_{4}X_{9}$ & $X_{4}X_{12}$ & $X_{1}X_{2}X_{4}$ & $X_{5}$ & $X_{4}X_{7}X_{8}$ & $X_{1}X_{4}X_{7}X_{10}$ & $X_{1}X_{4}X_{7}X_{11}$
\\ 
\hline 
\end{tabular} 
}

\newcommand{\ZIZsorted}{\begin{tabular}{|c| c c c c c c c c c c|}
\hline
000 & $ $ & $X_{2}$ & $X_{5}$ & $X_{8}$ & $X_{11}$ & $X_{1}X_{3}$ & $X_{4}X_{6}$ & $X_{7}X_{9}$ & $X_{1}X_{7}X_{10}$ & $X_{1}X_{7}X_{12}$
\\ 
101 & $X_{1}$ & $X_{1}X_{2}$ & $X_{1}X_{5}$ & $X_{1}X_{8}$ & $X_{1}X_{11}$ & $X_{3}$ & $X_{1}X_{4}X_{6}$ & $X_{1}X_{7}X_{9}$ & $X_{7}X_{10}$ & $X_{7}X_{12}$
\\ 
110 & $X_{7}$ & $X_{2}X_{7}$ & $X_{5}X_{7}$ & $X_{7}X_{8}$ & $X_{7}X_{11}$ & $X_{1}X_{3}X_{7}$ & $X_{4}X_{6}X_{7}$ & $X_{9}$ & $X_{1}X_{10}$ & $X_{1}X_{12}$
\\ 
011 & $X_{10}$ & $X_{2}X_{10}$ & $X_{5}X_{10}$ & $X_{8}X_{10}$ & $X_{10}X_{11}$ & $X_{1}X_{3}X_{10}$ & $X_{4}X_{6}X_{10}$ & $X_{7}X_{9}X_{10}$ & $X_{1}X_{7}$ & $X_{1}X_{7}X_{10}X_{12}$
\\ 
111 & $X_{4}$ & $X_{2}X_{4}$ & $X_{4}X_{5}$ & $X_{4}X_{8}$ & $X_{4}X_{11}$ & $X_{1}X_{3}X_{4}$ & $X_{6}$ & $X_{4}X_{7}X_{9}$ & $X_{1}X_{4}X_{7}X_{10}$ & $X_{1}X_{4}X_{7}X_{12}$
\\ 
\hline 
\end{tabular} 
}

\newcommand{\hammingbitflipZIZ}{\begin{tikzcd}[ampersand replacement=\&,row sep=0.05cm,column sep=0.10cm]
\lstick[wires=3]{\ket{\psi_1}}\&\qw\&\qw\error{X}{two_Set1}\&\ctrl{12} \&\qw\&\qw\&\qw\&\qw\&\qw\&\qw\&\qw\&\qw\&\qw\&\qw\&\qw\&\ctrl{14} \&\qw\&\qw\&\qw\&\qw\&\qw\&\qw\&\error{X}{white}\&\error{X}{white}\&\error{X}{white} \\
\&\qw\&\qw\error{X}{zero_Set1}\&\qw\&\qw\&\qw\&\qw\&\qw\&\qw\&\qw\&\qw\&\qw\&\qw\&\qw\&\qw\&\qw\&\qw\&\qw\&\qw\&\qw\&\qw\&\qw\&\error{X}{white}\&\error{X}{white}\&\error{X}{white} \\
\&\qw\&\qw\&\qw\&\ctrl{10} \&\qw\&\qw\&\qw\&\qw\&\qw\&\qw\&\qw\&\qw\&\qw\&\qw\&\qw\&\ctrl{12} \&\qw\&\qw\&\qw\&\qw\&\qw\&\error{X}{white}\&\error{X}{white}\&\error{X}{white} \\ [0.250000 cm]
\lstick[wires=3]{\ket{\psi_2}}\&\qw\&\qw\&\qw\&\qw\&\ctrl{9} \&\qw\&\qw\&\qw\&\ctrl{10} \&\qw\&\qw\&\qw\&\qw\&\qw\&\qw\&\qw\&\ctrl{11} \&\qw\&\qw\&\qw\&\qw\&\error{X}{white}\&\error{X}{white}\&\error{X}{white} \\
\&\qw\&\qw\&\qw\&\qw\&\qw\&\qw\&\qw\&\qw\&\qw\&\qw\&\qw\&\qw\&\qw\&\qw\&\qw\&\qw\&\qw\&\qw\&\qw\&\qw\&\qw\&\error{X}{white}\&\error{X}{white}\&\error{X}{white} \\
\&\qw\&\qw\&\qw\&\qw\&\qw\&\ctrl{7} \&\qw\&\qw\&\qw\&\ctrl{8} \&\qw\&\qw\&\qw\&\qw\&\qw\&\qw\&\qw\&\ctrl{9} \&\qw\&\qw\&\qw\&\error{X}{white}\&\error{X}{white}\&\error{X}{white} \\ [0.250000 cm]
\lstick[wires=3]{\ket{\psi_3}}\&\qw\&\qw\error{X}{one_Set1}\&\qw\&\qw\&\qw\&\qw\&\ctrl{6} \&\qw\&\qw\&\qw\&\ctrl{7} \&\qw\&\qw\&\qw\&\qw\&\qw\&\qw\&\qw\&\qw\&\qw\&\qw\&\error{X}{white}\&\error{X}{white}\&\error{X}{white} \\
\&\qw\&\qw\&\qw\&\qw\&\qw\&\qw\&\qw\&\qw\&\qw\&\qw\&\qw\&\qw\&\qw\&\qw\&\qw\&\qw\&\qw\&\qw\&\qw\&\qw\&\qw\&\error{X}{white}\&\error{X}{white}\&\error{X}{white} \\
\&\qw\&\qw\&\qw\&\qw\&\qw\&\qw\&\qw\&\ctrl{4} \&\qw\&\qw\&\qw\&\ctrl{5} \&\qw\&\qw\&\qw\&\qw\&\qw\&\qw\&\qw\&\qw\&\qw\&\error{X}{white}\&\error{X}{white}\&\error{X}{white} \\ [0.250000 cm]
\lstick[wires=3]{\ket{\psi_4}}\&\qw\&\qw\error{X}{two_Set1}\&\qw\&\qw\&\qw\&\qw\&\qw\&\qw\&\qw\&\qw\&\qw\&\qw\&\ctrl{4} \&\qw\&\qw\&\qw\&\qw\&\qw\&\ctrl{5} \&\qw\&\qw\&\error{X}{white}\&\error{X}{white}\&\error{X}{white} \\
\&\qw\&\qw\&\qw\&\qw\&\qw\&\qw\&\qw\&\qw\&\qw\&\qw\&\qw\&\qw\&\qw\&\qw\&\qw\&\qw\&\qw\&\qw\&\qw\&\qw\&\qw\&\error{X}{white}\&\error{X}{white}\&\error{X}{white} \\
\&\qw\&\qw\&\qw\&\qw\&\qw\&\qw\&\qw\&\qw\&\qw\&\qw\&\qw\&\qw\&\qw\&\ctrl{2} \&\qw\&\qw\&\qw\&\qw\&\qw\&\ctrl{3} \&\qw\&\error{X}{white}\&\error{X}{white}\&\error{X}{white} \\ [0.250000 cm]
\&\qw\&\qw\&\targ\qw \&\targ\qw \&\targ\qw \&\targ\qw \&\targ\qw \&\targ\qw \&\qw\&\qw\&\qw\&\qw\&\qw\&\qw\&\qw\&\qw\&\qw\&\qw\&\qw\&\qw\&\qw\&\error{X}{white}\&\error{X}{one_Set1}\&\error{X}{two_Set1} \\
\&\qw\&\qw\&\qw\&\qw\&\qw\&\qw\&\qw\&\qw\&\targ\qw \&\targ\qw \&\targ\qw \&\targ\qw \&\targ\qw \&\targ\qw \&\qw\&\qw\&\qw\&\qw\&\qw\&\qw\&\qw\&\error{X}{white}\&\error{X}{one_Set1}\&\error{X}{two_Set1} \\
\&\qw\&\qw\&\qw\&\qw\&\qw\&\qw\&\qw\&\qw\&\qw\&\qw\&\qw\&\qw\&\qw\&\qw\&\targ\qw \&\targ\qw \&\targ\qw \&\targ\qw \&\targ\qw \&\targ\qw \&\qw\&\error{X}{white}\&\error{X}{white}\&\error{X}{white} \\ [0.250000 cm]
\end{tikzcd}}

\newcommand{\hammingbitflipZZI}{\begin{tikzcd}[ampersand replacement=\&,row sep=0.05cm,column sep=0.10cm]
\lstick[wires=3]{\ket{\psi_1}}\&\qw\&\qw\error{X}{two_Set1}\&\ctrl{12} \&\qw\&\qw\&\qw\&\qw\&\qw\&\qw\&\qw\&\qw\&\qw\&\qw\&\qw\&\ctrl{14} \&\qw\&\qw\&\qw\&\qw\&\qw\&\qw\&\error{X}{white}\&\error{X}{white}\&\error{X}{white} \\
\&\qw\&\qw\error{X}{zero_Set1}\&\qw\&\ctrl{11} \&\qw\&\qw\&\qw\&\qw\&\qw\&\qw\&\qw\&\qw\&\qw\&\qw\&\qw\&\ctrl{13} \&\qw\&\qw\&\qw\&\qw\&\qw\&\error{X}{white}\&\error{X}{white}\&\error{X}{white} \\
\&\qw\&\qw\&\qw\&\qw\&\qw\&\qw\&\qw\&\qw\&\qw\&\qw\&\qw\&\qw\&\qw\&\qw\&\qw\&\qw\&\qw\&\qw\&\qw\&\qw\&\qw\&\error{X}{white}\&\error{X}{white}\&\error{X}{white} \\ [0.250000 cm]
\lstick[wires=3]{\ket{\psi_2}}\&\qw\&\qw\&\qw\&\qw\&\ctrl{9} \&\qw\&\qw\&\qw\&\ctrl{10} \&\qw\&\qw\&\qw\&\qw\&\qw\&\qw\&\qw\&\ctrl{11} \&\qw\&\qw\&\qw\&\qw\&\error{X}{white}\&\error{X}{white}\&\error{X}{white} \\
\&\qw\&\qw\&\qw\&\qw\&\qw\&\ctrl{8} \&\qw\&\qw\&\qw\&\ctrl{9} \&\qw\&\qw\&\qw\&\qw\&\qw\&\qw\&\qw\&\ctrl{10} \&\qw\&\qw\&\qw\&\error{X}{white}\&\error{X}{white}\&\error{X}{white} \\
\&\qw\&\qw\&\qw\&\qw\&\qw\&\qw\&\qw\&\qw\&\qw\&\qw\&\qw\&\qw\&\qw\&\qw\&\qw\&\qw\&\qw\&\qw\&\qw\&\qw\&\qw\&\error{X}{white}\&\error{X}{white}\&\error{X}{white} \\ [0.250000 cm]
\lstick[wires=3]{\ket{\psi_3}}\&\qw\&\qw\error{X}{one_Set1}\&\qw\&\qw\&\qw\&\qw\&\ctrl{6} \&\qw\&\qw\&\qw\&\ctrl{7} \&\qw\&\qw\&\qw\&\qw\&\qw\&\qw\&\qw\&\qw\&\qw\&\qw\&\error{X}{white}\&\error{X}{white}\&\error{X}{white} \\
\&\qw\&\qw\&\qw\&\qw\&\qw\&\qw\&\qw\&\ctrl{5} \&\qw\&\qw\&\qw\&\ctrl{6} \&\qw\&\qw\&\qw\&\qw\&\qw\&\qw\&\qw\&\qw\&\qw\&\error{X}{white}\&\error{X}{white}\&\error{X}{white} \\
\&\qw\&\qw\&\qw\&\qw\&\qw\&\qw\&\qw\&\qw\&\qw\&\qw\&\qw\&\qw\&\qw\&\qw\&\qw\&\qw\&\qw\&\qw\&\qw\&\qw\&\qw\&\error{X}{white}\&\error{X}{white}\&\error{X}{white} \\ [0.250000 cm]
\lstick[wires=3]{\ket{\psi_4}}\&\qw\&\qw\error{X}{two_Set1}\&\qw\&\qw\&\qw\&\qw\&\qw\&\qw\&\qw\&\qw\&\qw\&\qw\&\ctrl{4} \&\qw\&\qw\&\qw\&\qw\&\qw\&\ctrl{5} \&\qw\&\qw\&\error{X}{white}\&\error{X}{white}\&\error{X}{white} \\
\&\qw\&\qw\&\qw\&\qw\&\qw\&\qw\&\qw\&\qw\&\qw\&\qw\&\qw\&\qw\&\qw\&\ctrl{3} \&\qw\&\qw\&\qw\&\qw\&\qw\&\ctrl{4} \&\qw\&\error{X}{white}\&\error{X}{white}\&\error{X}{white} \\
\&\qw\&\qw\&\qw\&\qw\&\qw\&\qw\&\qw\&\qw\&\qw\&\qw\&\qw\&\qw\&\qw\&\qw\&\qw\&\qw\&\qw\&\qw\&\qw\&\qw\&\qw\&\error{X}{white}\&\error{X}{white}\&\error{X}{white} \\ [0.250000 cm]
\&\qw\&\qw\&\targ\qw \&\targ\qw \&\targ\qw \&\targ\qw \&\targ\qw \&\targ\qw \&\qw\&\qw\&\qw\&\qw\&\qw\&\qw\&\qw\&\qw\&\qw\&\qw\&\qw\&\qw\&\qw\&\error{X}{zero_Set1}\&\error{X}{one_Set1}\&\error{X}{two_Set1} \\
\&\qw\&\qw\&\qw\&\qw\&\qw\&\qw\&\qw\&\qw\&\targ\qw \&\targ\qw \&\targ\qw \&\targ\qw \&\targ\qw \&\targ\qw \&\qw\&\qw\&\qw\&\qw\&\qw\&\qw\&\qw\&\error{X}{white}\&\error{X}{one_Set1}\&\error{X}{two_Set1} \\
\&\qw\&\qw\&\qw\&\qw\&\qw\&\qw\&\qw\&\qw\&\qw\&\qw\&\qw\&\qw\&\qw\&\qw\&\targ\qw \&\targ\qw \&\targ\qw \&\targ\qw \&\targ\qw \&\targ\qw \&\qw\&\error{X}{zero_Set1}\&\error{X}{white}\&\error{X}{white} \\ [0.250000 cm]
\end{tikzcd}}

\newcommand{\hammingbitflipzeroFT}{\begin{tikzcd}[ampersand replacement=\&,row sep=0.05cm,column sep=0.10cm]
\lstick[wires=3]{\ket{\psi_{1}}}\&\ctrl{12} \&\qw\&\qw\&\qw\&\qw\&\qw\&\qw\&\qw\&\qw\&\qw\&\qw\&\qw\&\ctrl{16} \&\qw\&\qw\&\qw\&\qw\&\qw\&\qw \\
\&\qw\&\qw\&\qw\&\ctrl{12} \&\qw\&\qw\&\qw\&\qw\&\qw\&\qw\&\qw\&\qw\&\qw\&\qw\&\qw\&\ctrl{16} \&\qw\&\qw\&\qw \\
\&\qw\&\qw\&\qw\&\qw\&\qw\&\qw\&\qw\&\qw\&\qw\&\qw\&\qw\&\qw\&\qw\&\qw\&\qw\&\qw\&\qw\&\qw\&\qw \\ [0.250000 cm]
\lstick[wires=3]{\ket{\psi_{2}}}\&\qw\&\ctrl{9} \&\qw\&\qw\&\qw\&\qw\&\ctrl{11} \&\qw\&\qw\&\qw\&\qw\&\qw\&\qw\&\ctrl{13} \&\qw\&\qw\&\qw\&\qw\&\qw \\
\&\qw\&\qw\&\qw\&\qw\&\ctrl{9} \&\qw\&\qw\&\qw\&\qw\&\ctrl{11} \&\qw\&\qw\&\qw\&\qw\&\qw\&\qw\&\ctrl{13} \&\qw\&\qw \\
\&\qw\&\qw\&\qw\&\qw\&\qw\&\qw\&\qw\&\qw\&\qw\&\qw\&\qw\&\qw\&\qw\&\qw\&\qw\&\qw\&\qw\&\qw\&\qw \\ [0.250000 cm]
\lstick[wires=3]{\ket{\psi_{3}}}\&\qw\&\qw\&\ctrl{6} \&\qw\&\qw\&\qw\&\qw\&\ctrl{8} \&\qw\&\qw\&\qw\&\qw\&\qw\&\qw\&\qw\&\qw\&\qw\&\qw\&\qw \\
\&\qw\&\qw\&\qw\&\qw\&\qw\&\ctrl{6} \&\qw\&\qw\&\qw\&\qw\&\ctrl{8} \&\qw\&\qw\&\qw\&\qw\&\qw\&\qw\&\qw\&\qw \\
\&\qw\&\qw\&\qw\&\qw\&\qw\&\qw\&\qw\&\qw\&\qw\&\qw\&\qw\&\qw\&\qw\&\qw\&\qw\&\qw\&\qw\&\qw\&\qw \\ [0.250000 cm]
\lstick[wires=3]{\ket{\psi_{4}}}\&\qw\&\qw\&\qw\&\qw\&\qw\&\qw\&\qw\&\qw\&\ctrl{5} \&\qw\&\qw\&\qw\&\qw\&\qw\&\ctrl{7} \&\qw\&\qw\&\qw\&\qw \\
\&\qw\&\qw\&\qw\&\qw\&\qw\&\qw\&\qw\&\qw\&\qw\&\qw\&\qw\&\ctrl{5} \&\qw\&\qw\&\qw\&\qw\&\qw\&\ctrl{7} \&\qw \\
\&\qw\&\qw\&\qw\&\qw\&\qw\&\qw\&\qw\&\qw\&\qw\&\qw\&\qw\&\qw\&\qw\&\qw\&\qw\&\qw\&\qw\&\qw\&\qw \\ [0.250000 cm]
\lstick[wires=2]{\ket{\alpha_{1}}}\&\targ\qw \&\targ\qw \&\targ\qw \&\qw\&\qw\&\qw\&\qw\&\qw\&\qw\&\qw\&\qw\&\qw\&\qw\&\qw\&\qw\&\qw\&\qw\&\qw\&\qw \\
\&\qw\&\qw\&\qw\&\targ\qw \&\targ\qw \&\targ\qw \&\qw\&\qw\&\qw\&\qw\&\qw\&\qw\&\qw\&\qw\&\qw\&\qw\&\qw\&\qw\&\qw \\
\lstick[wires=2]{\ket{\alpha_{2}}}\&\qw\&\qw\&\qw\&\qw\&\qw\&\qw\&\targ\qw \&\targ\qw \&\targ\qw \&\qw\&\qw\&\qw\&\qw\&\qw\&\qw\&\qw\&\qw\&\qw\&\qw \\ 
\&\qw\&\qw\&\qw\&\qw\&\qw\&\qw\&\qw\&\qw\&\qw\&\targ\qw \&\targ\qw \&\targ\qw \&\qw\&\qw\&\qw\&\qw\&\qw\&\qw\&\qw \\
\lstick[wires=2]{\ket{\alpha_{3}}}\&\qw\&\qw\&\qw\&\qw\&\qw\&\qw\&\qw\&\qw\&\qw\&\qw\&\qw\&\qw\&\targ\qw \&\targ\qw \&\targ\qw \&\qw\&\qw\&\qw\&\qw \\
\&\qw\&\qw\&\qw\&\qw\&\qw\&\qw\&\qw\&\qw\&\qw\&\qw\&\qw\&\qw\&\qw\&\qw\&\qw\&\targ\qw \&\targ\qw \&\targ\qw \&\qw \\ 
\end{tikzcd}}

\begin{document}

\title{Quantum Error  Source and Channel Coding}
\author{Dennis Lucarelli}
\affiliation{Error Corp.}
\begin{abstract}
Quantum error correction draws many of its principles and constructions from classical coding theory, adapted to the unique aspects of quantum mechanics. Here we extend this correspondence to a block of logical qubits, each prepared in the same quantum code, by coupling the block through a classical error correcting code: syndrome qubits then carry error information from several logical qubits at once, and collective inference recovers the errors. Algebraically, the construction defines subgroups of the product stabilizer group corresponding to duals of classical codes, and we prove that a lookup table decoder corrects every error pattern respecting the correction radii of the two constituent codes. For classical algebraic code families, the number of syndrome qubits serving $L$ logical qubits scales as ${\cal O}(\log_2(L+1))$ at the code-capacity level, though at the price of stabilizer weight growing with the block length. An entropy bound shows that check weights need not grow with $L$. Under a phenomenological noise model, the same lookup table identifies measurement errors by nearest-neighbor post-processing, and classical algebraic decoders locate exactly the logical qubits carrying errors even with noisy syndromes. More broadly, we argue that quantum error correction, reduced to its functional core, is source compression in the sense of Shannon, whose source and channel coding theorems bound the overhead rates of quantum post-selection tasks.
\end{abstract}
\maketitle

\section{Introduction}
Quantum algorithms are known to achieve super-polynomial speedup in solving certain
problems in discrete mathematics or as universal simulators of quantum systems \cite{quantum_zoo}.  
Due to the fragility of quantum states, it is widely accepted that quantum error correction 
will be required to maintain quantum coherence for sufficiently long algorithm runtimes. However, 
circuit implementations of quantum algorithms of practical interest require billions, perhaps 
trillions, of time-steps. These so-called {\it deep}  circuits necessitate a practically flawless
error corrected or {\it logical}  qubit---one that fails less than
the reciprocal of the problem size.  For a quantum algorithm with $\numL$ logical qubits executing a
depth $\Delta$ circuit, we require the probability of a logical qubit error $P_L < 1/\Delta \numL$.
In principle, provided the underlying error rate is below a
certain threshold, a logical qubit of arbitrarily high fidelity can be constructed
by concatenating quantum codes \cite{knill1998resilient,10.1145/258533.258579}.
However, code concatenation leads to an explosion of quantum resources (i.e. number of qubits)
required to hold the data and the ancillary qubits needed for syndrome
extraction and error correction.  Resource studies suggest that an error corrected
quantum computer will be very large, requiring millions of qubits
\cite{reiher2017elucidating,babbush2018encoding,gidney2019factor},
although improved codes, algorithms, and compilations continue to
reduce these estimates \cite{gidney2025factor,bravyi2024high}.  Even so, such
machines remain well beyond present-day systems, and reducing the qubit
overhead of quantum error correction is a central challenge on the path to
fault-tolerant quantum computation.
Here, we show that qubit overhead can be significantly 
reduced by taking the  {\it gestalt} view of quantum error correction (QEC). We construct a QEC 
protocol defined across the entire quantum computer and show that ancillary or {\it syndrome}
qubits can carry error information from multiple  logical qubits simultaneously.  
From the syndrome measurements, collective inference can efficiently and provably
reconstruct the errors unambiguously, even in the presence of
measurement errors.
Underlying the construction is the observation that, at its functional core,
quantum error correction is not channel coding but source coding: the decoder
never receives a message; it receives only the syndrome, a compression of an
unknown error pattern.
We state the accompanying trade-off at the outset: the compression comes at
the cost of stabilizer weight. The weight of a product stabilizer is the
product of the weights of its classical and quantum factors, so each syndrome
qubit couples to data qubits across multiple logical qubits, increasing the
number of two-qubit gates and the connectivity required per syndrome
measurement. We quantify this cost in Sec.~\ref{noisy_encoding}, address
fault-tolerant implementation in Sec.~\ref{fault_tolerance}, and show in
Sec.~\ref{ldpc_bounds} that with low-density classical codes the check
weights need not grow with the block length.

In the context of quantum communication over a Pauli channel, 
the {\it hashing bound}  \cite{PhysRevA.54.3824} characterizes the 
achievable rate  of transmitting $k$ message qubits
encoded in $n$ physical qubits as asymptotically limited by the Shannon entropy of
the channel,  $ k/n < 1 - H_2({\mathbf p}).$ Here,
${\mathbf p} = \left[ \, p_x, p_y, p_z \, \right]$ is a vector of Pauli channel error probabilities
and $H_2({\mathbf p})$ denotes Shannon's binary entropy function.
This bound can be seen as a special case of the  quantum channel capacity 
\cite{lloyd1997capacity,LSD_shor,LSD_devetak}. Moreover, using Shannon's
methods of typicality and random code constructions, quantum stabilizer codes
 \cite{gottesman1997stabilizer} 
can achieve the quantum channel capacity with a Pauli error model for 
large enough $n$ \cite{wilde2013quantum}.  Strictly speaking,  however, 
QEC is not channel coding, as typically construed in the classical theory. In 
channel coding, a message's error syndrome is computed at the 
encoder, appended to the message and sent over a noisy channel.  At the receiver, the 
decoder has access and utilizes {\it both} the noise corrupted message and the syndrome to
reconstruct the message. In QEC---as opposed to quantum communication---the decoder has
access to the error syndrome only.

Up to an irrelevant phase, the Pauli group on $n$ qubits 
$\Pauli_n = \langle \, \pm i I, X,Y, Z\,\rangle^{\otimes n}$ 
is isomorphic to the binary vector space ${\mathbb F}_2^{2n}$
with modulo 2 arithmetic and the symplectic inner product \cite{calderbank1997quantum}.
By this correspondence, syndrome extraction is equivalent to mapping a length $2n$ bit string
(accounting for both $X$ and $Z$-type errors) to a length $n-k$ bit syndrome, and
thus performs compression of {\it classical} data---albeit
{\it unknown} classical data derived from  quantum data. In the classical
literature this primitive is known as {\it syndrome source coding}
\cite{ancheta1976syndrome}. This is the operational 
meaning of the hashing bound, which can be restated 
as Shannon's source coding theorem \cite{shannon1948mathematical} for linear block codes
bounding the optimal compression rate of binary data by the entropy of the random source, namely
$n-k >  n H_2({\mathbf p}).$ Here, ${\mathbf p}$ parameterizes a Pauli {\it error source}.
The rate statement is well known, though seldom framed as such in the quantum
literature. Indeed, the correspondence is sharper than an analogy: in QEC the
decoder expects the {\it zero message} (no errors), so nothing is
communicated, and a channel code carrying a known message is precisely a
source code for the noise. Reduced to its functional core, quantum error
correction is source coding of the error process, and we adopt this view
throughout. Our contribution is to take this view seriously as an
architecture. Shannon's bounds hold for blocks of logical qubits, syndrome
extraction may be pooled across the register, and classical source and
channel codes, together with their decoders, apply directly.

Given an error syndrome $y$ and a linear error correcting code specified by the 
parity-check matrix $H,$ the maximum-likelihood decoding problem is to determine
the vector $x$ with a minimum number of non-zero entries such that $Hx = y.$ This inverse problem 
is known to be NP-hard for classical linear error correcting codes \cite{berlekamp1978hca}, implying 
that an efficient maximum-likelihood decoder is not likely to exist. Similar complexity arguments have been shown 
for quantum decoders \cite{hsieh2011np, iyer2015hardness}. Further complicating the quantum decoding problem 
is that errors in the syndrome measurements are almost always present.  This is an 
unavoidable consequence of faulty two-qubit gates and quantum measurements, which often have error 
rates orders of magnitude higher than data qubit errors.  
A number of methods have been proposed to account
for syndrome measurement errors in quantum decoding, including minimum 
weight perfect matching decoders \cite{dennis2002topological,PhysRevLett.108.180501}
and iterative belief-propagation based decoders \cite{PhysRevLett.122.200501}.  
Here, we construct a lookup table decoder, designed for a restricted class of error patterns acting 
on a block of logical qubits, that can be queried with nearest-neighbor binary string matching to 
unambiguously determine the errors in the data qubits while also protecting 
against errors in the syndrome measurements.   

Quantum product code constructions have been proposed previously in the literature.
In \cite{grassl2005quantum}, using the CSS construction, conditions were
derived for obtaining a quantum error correcting code from
the (direct) product of two classical error correcting codes.
In \cite{10.5555/3179575.3179578},  quantum tensor
product codes, the quantum analogues of  generalized tensor product codes
\cite{wolf1965codes,imai1981generalized}, where the constituent codes
are defined over a field and an extension field, were investigated.
Notably,  quantum hypergraph-product codes \cite{tillich2013quantum} employ
the Kronecker product to construct a quantum parity-check matrix from two
classical codes.  Product constructions have since become central to quantum
coding theory: generalizations of the hypergraph product, including balanced
products \cite{breuckmann2021balanced} and lifted products
\cite{panteleev2022asymptotically}, yield quantum low-density parity-check
(LDPC) codes of constant rate and linear distance
\cite{panteleev2022asymptotically,leverrier2022tanner}
(see \cite{breuckmann2021qldpc} for a review).
These  product constructions encode multiple logical
qubits in a single block of data qubits and define proper quantum error correcting codes.
In contrast, as discussed in the following, the method
proposed here, while based on a product code construction,
is not strictly a quantum code. Rather, we propose the
use of any binary, classical code linking  multiple copies of a
single quantum code.

Notation and background on classical and quantum error correcting 
codes may be  found in Appendix \ref{appback}.

\section{The Product Construction}\label{results}
We propose a method that 
detects and locates errors in a length-$\numL$ block of 
logical qubits by constructing stabilizer circuits
acting on the entire block of logical qubits and collectively 
processing the syndrome measurement results.  In particular, 
we construct stabilizers in the product group
\be
\stabblock\ \definedby  \stabgroup_1 \times \stabgroup_2  \, \times \,  
\cdots \,  \times \, \stabgroup_{\numL}
\label{stabblock}
\ee
where $ \stabgroup_{\ell}$ 
is the stabilizer group of a quantum error correcting code
$\qcode$ acting on logical qubit $\ell.$ Let  $\qcode \sim \QEC{n}{k}{\qd}$
denote a quantum code encoding $k$ logical qubits in $n$ physical qubits
with code distance $\qd,$ and capable of correcting $\qtt = \lfloor \frac{\qd -1}{2} \rfloor $
errors.  The construction is valid for any $\qk$: when $\qk > 1,$ each factor
of (\ref{stabblock}) protects a block of $\qk$ logical qubits, and the results
that follow hold with logical qubits replaced by code blocks. As all examples
considered in this work employ $\qk = 1$ codes, we refer to each encoded block
as a logical qubit throughout; for $\qk > 1,$ the failure probabilities of
Sec.~\ref{failure_prob} are likewise per-block quantities.
Defining the total number of data qubits as $\nn = \numQ \numL,$ we have
$\stabblock \subset \Pauli_N.$

Suppose we have a block of $\numL=7$ logical qubits.  
For  arbitrary $\stab \in \stabgroup$ and the identity $e,$ 
consider the following elements of $\stabblock$  
\begin{eqnarray}
\stab\,  e\,        \stab	\,e\,	   \stab\,	e\,	\stab  \nonumber \\
e	\, \stab\,  \stab	\,e\,	   e\, 	     \stab\,  \stab\\
e	\, e	   \, e\,	\stab\, \stab\,  \stab\,    \stab \nonumber
\end{eqnarray} 
One may recognize the stabilizer elements $\left\{ e, \, \stab \right\}$  
arranged (columnwise) in the binary representation of the 
integers one through seven or, equivalently, 
according to the parity-check matrix of a classical 
Hamming $\left[ 7, 4,3 \right]$ error correcting code.   
The key observation 
is that  this ``encoding" is realized by the Kronecker product between a
binary matrix and a parity-check matrix of a quantum error correcting code. 
In particular, the Pauli operator corresponding to the Kronecker product of any binary vector with the 
binary representation of a stabilizer generator is again
in the product stabilizer $\stabblock$. The ordering of the product is critical---post-multiplying
by a binary matrix does not produce an element of the stabilizer.
As commuting elements in  $\stabblock,$  
measurements of Pauli operators corresponding
to rows of a matrix constructed by pre-multiplying a 
stabilizer by a binary matrix  can be performed 
without leaving the codespace and destroying the 
quantum information held in the data qubits.

The canonical scheme for syndrome extraction from $\numL$ logical qubits, 
where separate blocks of syndrome qubits are allocated to each logical qubit
and each stabilizer, corresponds to a parity-check matrix 
$\id_\numL \,\kron \,\hq$ formed by the  parity-check matrix for $\qcode$ 
and the  $\numL \times \numL$ identity matrix.  In contrast, 
our construction first selects a classical error correcting code $\lcode$
with parameters $\lparams,$ assuming throughout that $\ld \geq \qd,$ and measures the stabilizers
corresponding to rows of the matrix formed by the  Kronecker product of a parity-check
matrix of the classical code $\lcode$ and the quantum code $\qcode $ 
\begin{equation} 
\hl\,\kron\,\hq
\label{hkronh}
\end{equation} 

The layout of data qubits for performing error correction depends on the 
choice of quantum code and device architecture.  Here, we conceptually arrange the data qubits
in a rectangular array with each column corresponding to the 
data qubits comprising a logical qubit. Under the
Pauli-to-binary isomorphism, 
unknown error patterns $\left [ \, \epX \,| \,\epZ \,\right]^\tp$ are then represented 
by $2\numQ \, \times \, \numL$ binary matrices with constituent bit-flip $\epX$ and 
phase-flip $\epZ$ matrices of dimension $\numQ \, \times \, \numL.$ Our construction applies to CSS codes 
\cite{calderbank1996good,steane1996multiple}, for which bit-flip and phase-flip corrections
can be performed independently, so, without loss of generality, we will denote an error pattern
by the binary matrix $\ep,$ which may 
represent $\epX$ or $\epZ.$ The number of errors in the data qubits is given by the 
Hamming weight $\wt{\ep},$ defined as the number of ones in the error pattern.  When applied to a
Pauli-$X$ or $Z$ operator, the Hamming weight is the number of non-identity operators.
The Hamming distance is the number of places that differ between 
two error patterns $\ep$ and $\ep^\prime$ and is equivalent to $\wt{\ep \oplus \ep^\prime}.$
 With a slight abuse of notation, the product ${\bf u} \otimes {\bf v}$ will denote both the Pauli operator 
obtained by performing the matrix Kronecker product applied to row vectors  and its 
equivalent binary rectangular representation ${\bf v}^\tp {\bf u}.$ 

 A general error pattern may be expressed as a sum of weight-1 error patterns 
\be
\ep =  \bigoplus_{\left(q , \ell\right)} \epE_{q\ell}
\label{error_pattern}
\ee
where $\epE_{q\ell}$ has a single non-zero element at the $\left(q,\ell\right)$-th entry
and the pairs $\left(q,\ell\right) \in \left(\idxset{\numQ} \, , \, \idxset{\numL} \right).$ 
The action of syndrome extraction performed according to the parity-check matrix
$\hl \otimes \hq$ can be represented by the matrix 
products defined by the Kronecker product identity
\be
\hl \otimes \hq \mathrm{vec}(\ep) = \mathrm{vec}(\hq \ep \hl^\tp) 
\label{vec_trick}
\ee
where $\mathrm{vec}(\ep)$ denotes the vector representation of an error pattern 
obtained by stacking the columns of $\ep$ and arithmetic is modulo 2. 
Suppose $\ep = \epE_{q\ell}$ is a weight-1 error pattern, then
\be
\hq \epE_{q\ell} = \left[ {\bf 0} \, \cdots \,  {\bf 0} \,  \, \syndQ_q \,  \, {\bf 0} \, \cdots {\bf 0} \right]
\label{hqE}
\ee
where the $\qcode${\it -syndrome} $\syndQ_q = \hq \epi{q}^\tp$ is the $\ell$-th column of the 
matrix on the right-hand side and $\epi{q}$ denotes a vector with a 1 in the $q$-th entry
and zeros elsewhere. The classical parity-check
matrix $\hl$ acts on the rows of $\hq \epE_{q\ell}.$  If the $\idxS$-th entry of
$\syndQ_q$ is 1, the $\lcode${\it -syndrome} $\syndL_\ell = \hl \epi{\ell}^\tp$ appears as the $\idxS$-th
row of $\hq \epE_{q\ell} \hl^\tp$ yielding
\be
\hq \epE_{q\ell} \hl^\tp =  \syndQ_q  \syndL_\ell^\tp 
\label{skrons}
\ee
The {\it product syndrome} $\syndCQ$ of a general error pattern $\ep,$ with $\wt{\ep}>1,$ 
is a binary sum over weight-1 syndromes
\be
\syndCQ =  \bigoplus_{\left(q,\ell\right)} \, \syndQ_q  \syndL_\ell^\tp
\ee

\begin{figure}[t!]
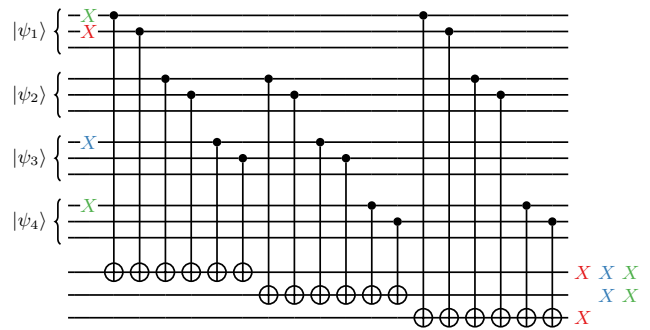

\begin{center}
\resizebox{0.475\textwidth}{!}{\hammingbitflipZZI}
\caption{Errors propagating through the circuit representation of $\pl^\tp  \otimes \left[ 1 \, 1\, 0\, \right]$
corresponding to the three-qubit bit-flip code stabilizer $\stab = ZZI.$ Four logical qubits 
$|\psi_\ell \rangle$ enter the circuit and are subject to bit-flip errors $(``X")$ highlighted
in color.  Using the propagation rules of Pauli-$X$ operators under CNOT gates,
errors propagate to the syndrome qubits as shown in the lower right portion
of the circuit.}
\label{ZZI_circuit}
\end{center}
\end{figure}

\subsection{Illustrative Example}\label{bitflip_example}
As a concrete example, consider a block of logical
qubits encoded in the three-qubit repetition code 
protecting against a single bit-flip acting on the codewords
\be
|{\bar 0\rangle} =  | 0 0 0 \rangle \quad \quad |{\bar 1\rangle} = |1 1 1 \rangle
\ee
The stabilizer group $\stabgroup = \langle ZZI, ZIZ \rangle$ is 
represented by the binary parity-check matrix 
\be
\hq =  \Hthreequbitmat 
\ee
The Pauli-$Z$ operator corresponding to the product 
$\left[ \, 1 \, 1 \, 0 \, 1 \, \right] \otimes   \left[ \, 1 \, 1 \, 0 \,\right]  $  is given by
$ZZIZZIIIIZZI$
which is an element of $\stabblock$ for the three-qubit bit-flip code.  A construction that 
encodes 7 logical qubits is based on the classical $\left[7,4,3\right] $ Hamming 
code with parity-check matrix in systematic form given by
\be
\left[ \, \id_3 \, \, | \, \, \pl^\tp \,\right] = \Hhamming
\label{A_mat}
\ee 
For brevity, we choose $\hl = \pl^\tp$ to encode 4 
logical qubits and construct the parity-check matrices 
\benonum
\pl^\tp \kron \left[ \,1 \, 1 \, 0\,\right] = \PkronZZI
\eenonum

\benonum
\pl^\tp \kron \left[ \,1 \, 0 \, 1\,\right]   = \PkronZIZ
\eenonum

\begin{figure}[t!]
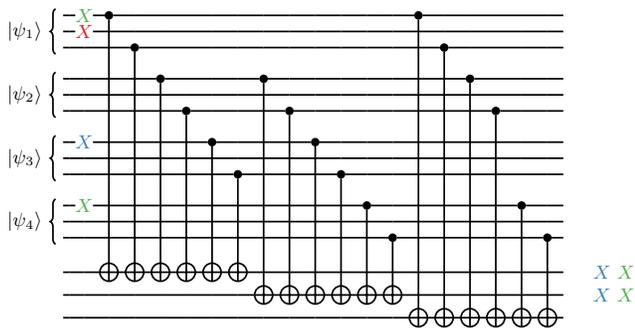

\begin{center}
\resizebox{0.4750\textwidth}{!}{\hammingbitflipZIZ}
\caption{Errors propagating through the circuit representation of $\pl^\tp  \otimes \left[ 1 \, 0\, 1\, \right]$
corresponding to the three-qubit bit-flip code stabilizer $\stab = ZIZ.$ Four logical qubits 
$|\psi_\ell \rangle$ enter the circuit and are subject to bit-flip errors $(``X")$ highlighted
in color.  Errors propagate to the syndrome qubits as shown in the lower right portion
of the circuit. Note that $ZIZ$ does not couple the second data qubit to the syndrome
qubits and, thus, the $X_2$ (red) does not propagate to the syndrome qubits. }
\label{ZIZ_circuit}
\end{center}
\end{figure}

Stabilizer circuits are constructed by coupling the data qubits  to the syndrome qubits  
as specified by a parity-check matrix in the usual way:  a two-qubit
gate is added to the circuit from data qubit $j$ to syndrome qubit $i$ 
if and only if the $(i,j)$-entry of the parity-check matrix is 1.  
Returning to the example, Figs. \ref{ZZI_circuit}  and \ref{ZIZ_circuit} 
show stabilizer circuits of $ZZI$ and $ZIZ$  encoded with $\pl^\tp.$
Four logical qubits $| \psi_\ell \rangle,$ encoded in the 
three-qubit bit-flip code, enter the circuit and are subject to errors.  There are 
a total of 12 data qubits (indexed from top to bottom) and 3 syndrome qubits per stabilizer. 
Using the propagation rules of Pauli-$X$ operators under CNOT gates, we may track the
errors through the circuit and observe how the parity relations defined by $\pl^\tp $
manifest in the syndrome qubits.  Let $E_j$ denote an $E$-type Pauli error on data qubit $j$. 
A bit-flip occurring on the second physical qubit $X_2$ (red $``X"$) is coupled to the first
and third syndrome qubits and a measurement in the $Z$ basis will yield the binary
outcomes $\left[\,1\,0\,1\,\right]$ (red $``X"$ on syndrome qubits one and three).  
Similarly, an error on $X_7$ (blue $``X"$) propagates to the first and 
second syndrome qubit as dictated by the seventh column of  $\pl^\tp \otimes \left[\,1\,1\,0\,\right]$. 
As in the classical case, there is ambiguity whenever the number of errors exceeds
the  correction radius $\qtt = 1,$  as evidenced by the
weight-2 error $X_{1}X_{10}$ (green) producing
syndrome measurement outcomes identical to the single qubit error $X_7$ (blue). The circuit 
couples the weight-2 error to the third syndrome qubit twice, negating 
the propagation of the error to that syndrome qubit. 

Syndrome ambiguity may be understood by examining the cosets of 
the Pauli-$X$ subgroup $\Pauli_{12} = \left\{ I, X \right\}^{\otimes 12}$
under the action of the Pauli-$Z$ operator corresponding to $ \pl^\tp \kron \left[\,1 \, 1 \, 0 \,\right].$ 
In the rectangular representation,  the syndrome conflict depicted in Fig. 
\ref{ZZI_circuit} is produced by the error patterns 
\be
X_7 \simeq \left[
\begin{array}{c c c c}
0&0&1&0\\
0&0&0&0\\
0&0&0&0\\
\end{array}
\right] \quad 
X_1X_{10} \simeq \left[
\begin{array}{c c c c}
1&0&0&1\\
0&0&0&0\\
0&0&0&0\\
\end{array}
\right] \quad 
\ee
Both error patterns are within the correction radius $\qtt=1$ for $\qcode$
(weight-1 errors on $|\psi_3\rangle$ and $\left\{ |\psi_1\rangle, |\psi_4\rangle\right\}$ 
respectively) but the weight-2 error $X_1X_{10}$ exceeds
the correction radius $\ltt=1$ of the classical code $\lcode.$ This results in 
identical first rows of ${\syndL}^\tp$ for both error patterns under the action of $\hl^\tp$
\be
 \left[
\begin{array}{c c c c}
0&0&1&0
\end{array}
\right] \pl
= 
 \left[
\begin{array}{c c c c}
1&0&0&1
\end{array}
\right] \pl = 
 \left[
\begin{array}{c c c}
1&1&0
\end{array}
\right]
\ee
In other words, the row vectors $\left[ \,0\,0\,1\,0 \,\right]$ and $\left[ \,1\,0\,0 \,1\,\right]$
are in the same coset of $\pl.$  The measured syndromes ``after" the action of 
the stabilizer $\left[ \,1 \,  1 \, 0 \, \right]$ will be identical and lead to a conflict.

A lookup table decoder exists for a class of errors $\errclass$ if there is a one-to-one mapping from
syndrome measurements to error patterns from that class. Some quantum codes are
degenerate, in the sense that multiple error patterns may result in the same syndrome measurement
but can all be corrected by a single Pauli correction operator. 
For degenerate codes,  the lookup table is modified to contain error syndrome-correction operator pairs.  
In either case, a lookup table may be pre-computed by iterating through all correctable 
error patterns $\ep \in \errclass$ and using the binary matrix relation $\hq \ep \hl^\tp$ to 
simulate the syndrome measurement outcomes.  To construct a lookup table decoder, we must
choose a single error operator from each coset, the {\it coset leader}, to represent 
the coset corresponding to a measured syndrome.  A table of syndrome-coset pairs 
with the cosets ordered with the coset representative in 
the first column is known as a {\em standard array}. Portions of the standard 
arrays corresponding to syndrome extraction performed according to
$\pl^\tp  \otimes \left[ \, 1\, 1\, 0 \,\right]$ and $\pl^\tp \otimes \left[\,1\,0\,1\,\right]$
are shown in Tables \ref{appbit}.\ref{syndtablesZZI} and
\ref{appbit}.\ref{syndtablesZIZ} in Appendix \ref{appbit}.
A lookup table decoder for single bit-flips may be constructed by joining these tables 
and defining the coset leader to  be the error pattern of minimum weight in the coset as 
shown in Table \ref{appbit}.\ref{example_lookup}.  When syndrome extraction for 
all stabilizers is performed in parallel via $\hl \, \otimes \, \hq$ each row of product 
syndrome $\syndCQ$ corresponds to a different stabilizer and the flattened product
syndrome as shown in Table \ref{appbit}.\ref{example_lookup} can be obtained by
performing $\mathrm{vec}(\syndCQ^\tp)$.
Assuming perfect CNOT gates and measurements, 
the lookup table decoder simply queries the keys of the table with the full syndrome measurement
and returns the unique data qubit corresponding to the matched key to be corrected.  

\subsection{Subgroups of $\stabblock$}\label{subgroups}
Generators of the product stabilizer group $\stabblock$ may be expressed in terms of Kronecker products
\be
\stabblock = \langle \, \epi{\ell} \otimes \stab\,  \rangle
\label{stabblock_compact}
\ee
Using the aforementioned isomorphism, we will interpret mixed products 
between binary vectors $\left(\epi{\ell}\in {\mathbb F}^\numL_2\right)$ and Pauli 
operators $\left( \stab \in \Pauli_\numQ \right)$ as well defined by implicitly mapping
Pauli operators to their binary representation, performing the matrix Kronecker product 
applied to row vectors, and mapping the result back to Pauli operators. 

A Pauli operator that commutes with the stabilizer group leaves the code space invariant
and is unobservable during syndrome extraction.  These operators form a normal subgroup of the
Pauli group generated by  $\numQ + \qk$ operators 
known as the {\it normalizer} $ \normS$  of $\stabgroup$ in $\Pauli_\numQ.$ 
Stabilizers and logical qubit rotations are in the normalizer as are non-trivial error patterns that go 
undetected during syndrome extraction.   A code's distance, defined as the minimal weight Pauli 
operator in $ \normS \setminus \stabgroup,$ 
is the distinguishing parameter of a quantum code, as it determines
the maximum weight of detectable and correctable errors. 
As such, one must characterize the structure of the normalizer 
to determine the error correcting capabilities of a quantum  code. 
The normalizer ${\cal N}(\stabblock)$ of the product group $\stabblock$ is generated by 
\be
{\cal N}(\stabblock) = \langle \, \epi{\ell} \otimes \eta_a \,\rangle
\ee
where $\eta_a \in \normS$ and $a \in \idxset{n+k}.$ 

In classical coding theory, the rows $\left\{ h_r \right\}$ of the parity-check matrix $\hl$ form 
a vector space $\lcode^\perp$ of dimension $\lred \definedby \numL - \lk$ known as the {\it dual code} of $\lcode.$  
For all $\stab \in \stabgroup$ and $r\in \idxset{R},$ define the group 
\be
\stabH \definedby \langle \, h_r \otimes \stab  \, \rangle
\ee
The following proposition establishes the structure of $\stabH$ as a
subgroup of $\stabblock.$

\begin{proposition}\label{subgroup_prop}
$\stabH$ is a subgroup of $\stabblock$ of rank $\lred \left( \numStab \right),$
whereas $\stabblock$ has rank $\numL \left( \numStab \right).$ Consequently, whenever
$\lk \geq 1,$ $\stabH$ is a proper subgroup of index
$2^{\lk \left( \numStab \right)}.$ Moreover, a generator $\epi{\ell} \otimes \stab$ of $\stabblock,$ with
$\stab$ not the identity, lies in $\stabH$ if and only if $\epi{\ell} \in \lcode^\perp.$
\end{proposition}

\begin{proof}
Containment follows by writing
$h \otimes \stab  = \prod_{\ell'} \epi{\ell'} \otimes \stab \in \stabblock,$ where $\ell'$
runs over the non-zero components of $h.$ Since $-I \notin \stabblock,$ the binary
representation is faithful and group multiplication is vector addition.
The generators of $\stabH$ are the elementary tensors
$h_r \otimes s_\idxS$ taken over spanning sets of $\lcode^\perp$ and of the span $S$ of
the binary representations of the stabilizer generators; hence the binary representation
of $\stabH$ is the tensor product space $\lcode^\perp \otimes\, S$ of dimension
$\lred \left( \numStab \right).$ Replacing $\lcode^\perp$ with ${\mathbb F}_2^{\numL}$
gives the corresponding statement for $\stabblock,$ of dimension
$\numL \left( \numStab \right),$ and the index follows. Finally, for non-zero $\stab,$ the
elementary tensor $\epi{\ell} \otimes \stab$ lies in $\lcode^\perp \otimes\, S$ if and only
if $\epi{\ell} \in \lcode^\perp.$
\end{proof}

The index measures the coarsening of syndrome information: the number of
distinguishable syndrome values falls by the factor $2^{\lk \left( \numStab \right)},$
that is, measuring generators of $\stabH$ in place of generators of
$\stabblock$ extracts $\lk \left( \numStab \right)$ fewer syndrome bits. Note that $\epi{\ell} \in \lcode^\perp$
occurs only if every codeword of $\lcode$ vanishes in the $\ell$-th coordinate, that is,
only if the dual code contains a weight-1 codeword. For any full-support code---including
the Hamming and BCH codes considered here---no generator $\epi{\ell} \otimes \stab$ of
$\stabblock$ lies in $\stabH.$
Group multiplication $\left( \,\odot \,\right)$ of generators  of $\stabH$ (and $\stabblock$)
in Kronecker form
is defined by applying the distributive property of the Kronecker product over multiplication
in $\Pauli_\numQ$ and binary addition in
${\mathbb F}_2^\numL$ to obtain
\be
h\otimes s \odot h'\otimes s'  = h \oplus h' \otimes ss' \cdot h \otimes s' \cdot h' \otimes s
\label{group-multiplication}
\ee
consistent with the addition of binary representations in the proof of
Proposition \ref{subgroup_prop}.
Up to a rearrangement of rows, $\stabH$ is the product  
subgroup corresponding to $\hl \, \kron \, \hq .$

As elements of $\Pauli_N,$ operators commute if and only if their binary
representations are orthogonal under the symplectic inner product.
For CSS codes, the symplectic inner product simplifies to  the modulo 2 inner product between
the binary representations of an $X$-type and a $Z$-type Pauli operator. 
Thus, the normalizer of the product subgroup 
$\stabH$ is the set of Pauli operators in the kernel of 
$\hl   \otimes \hq,$ or equivalently, binary matrices $\epN$ 
such that  $\hq \epN \hl^\tp = {\mathbf 0}.$ 
Let $\left\{g_m\right\}$ denote a basis of the classical code $\lcode$ 
satisfying $g_m \hl^\tp = {\mathbf 0}$ for $m \in \idxset{\lk}.$ 
The normalizer $\normH$ is generated by a set of {\it column generators}
$\epi{\ell} \otimes \eta_a,$ so called because $\eta_a$
appears in the $\ell$-th column in the rectangular representation,
and for  $E_q \in \left\{ X_q, Z_q \right\},$ a set of
{\it row generators} $g_m \otimes E_{q}$  where $g_m$ appears as
the $q$-th row in the rectangular representation.
Any product of row or column generators is an element of the 
normalizer as are cross products of the form $g \otimes \eta.$
Also in the normalizer are non-identity Pauli operators satisfying the conditions
\be
\hq \epN \subset {\cal C} \quad  \quad \epN \hl^\tp \subset \normS
\label{ep_conditions}
\ee
where the matrices are viewed as collections of rows (left) or columns (right). 

As a non-trivial example of this type of normalizer, consider a product code  formed by a   
block of 15 Steane  logical qubits \cite{steane1996multiple} 
 and the $\left[ 15,11,3 \right]$ 
classical Hamming code. A weight-9 normalizer element 
$\epN$ (say of $X$-type) may be illustrated as
\benonum
\includegraphics[width=0.25\textwidth]{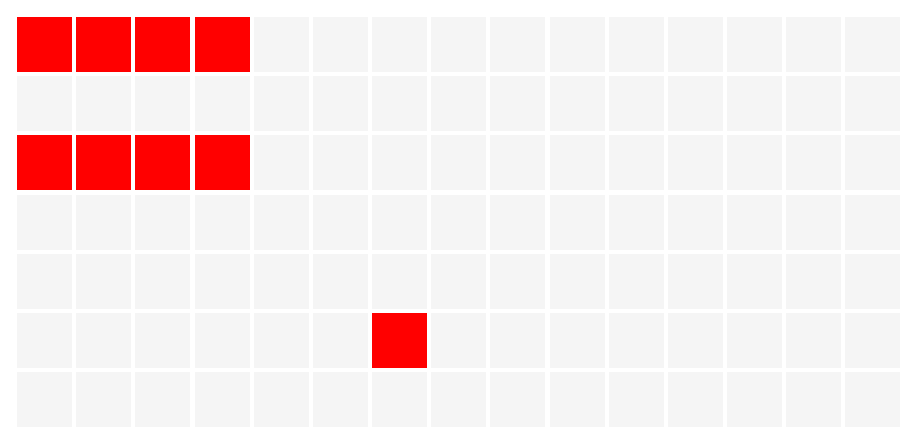}
\eenonum
where the red squares mark error locations. 
Under the action of $\hq, \, \epN$ transforms to
\benonum
\includegraphics[width=0.25\textwidth]{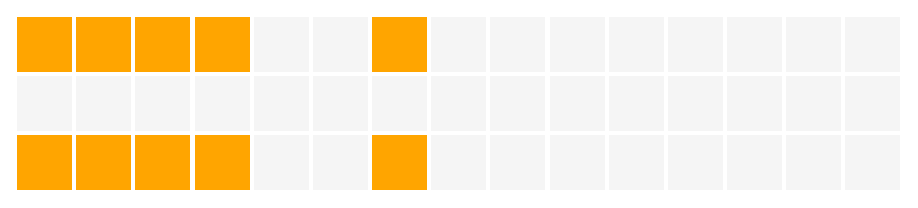}
\eenonum
from which it is evident that the non-zero columns of $\epN$ lie in the same coset of $\hq$ and 
map to identical columns of $\hq  \epN,$ producing rows in $\lcode.$ 
Similarly, $\epN$ transforms to
\benonum
\includegraphics[width=0.08\textwidth]{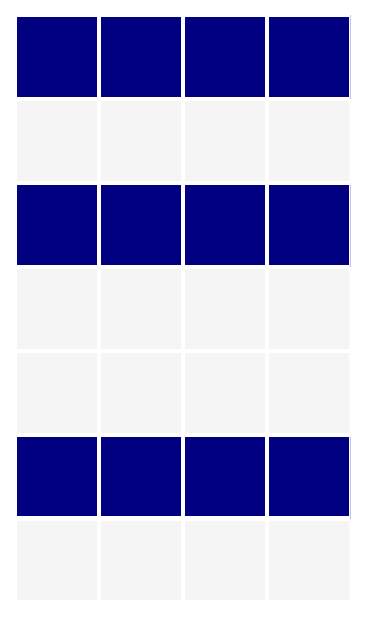}
\eenonum
with columns in ${\cal N}(\stabgroup)$ under the action of $\hl^\tp.$

Operators satisfying (\ref{ep_conditions}) can be written in terms of  row 
and column generators. The left-hand condition implies
\be
\hq \epN =  \bigoplus_{(m,\idxS)}g_m \otimes E_\idxS =  \bigoplus_{(m,\idxS)}g_m \otimes \hq {\mathbf v}_\idxS 
\ee
where $ \idxS \in \idxset{\numStab}$ 
and ${\mathbf v}_\idxS$ is such that $\hq {\mathbf v}_\idxS= E_\idxS.$ 
The right-hand condition implies 
 \be
\epN \hl^\tp =  \bigoplus_{(r,a)} \epi{r} \otimes \eta_a =  \bigoplus_{(r,a)} \hl {\mathbf u}_r \otimes \eta_a 
\ee
where $r \in \idxset{R}$ and 
${\mathbf u}_r$ is such that $\hl {\mathbf u}_r = \epi{r},$ collectively yielding  
\be
\epN =   \bigoplus_{(m,\idxS)} g_m \otimes {\mathbf v}_\idxS  \, \,   \bigoplus_{(r,a)} {\mathbf u}_r \otimes \eta_a
\label{ep_basis}
\ee
Summarizing, we have identified a dependent set of $\numL (n+k) + 2\lk\numQ$ operators
\be
\normH = \langle \,  \epi{\ell} \otimes \eta_a  \,, \,  g_m \otimes E_q \,\rangle
\label{NH}
\ee
That this set generates all of $\normH$ follows from a dimension count in the rectangular
representation. For each error type, let $\rho$ denote the rank of the quantum
parity-check matrix detecting that type, so that the kernel condition
$\hq \epN \hl^\tp = {\mathbf 0}$ defines a space of dimension
$\numQ \numL - \lred \, \rho.$ The column generators span the space of matrices whose
columns lie in the kernel of $\hq,$ of dimension $\numL \left( \numQ - \rho \right)$; the
row generators span the space of matrices whose rows lie in $\lcode,$ of dimension
$\lk \numQ$; and their intersection, spanned by the cross products $g \otimes \eta,$ has
dimension $\lk \left( \numQ - \rho \right).$ The three terms combine to
$\numL \left( \numQ - \rho \right) + \lk \numQ - \lk \left( \numQ - \rho \right) =
\numQ \numL - \lred \, \rho,$ matching the dimension of the kernel. We have thus
established the chain of subgroups
$\stabH \,\subset\, \stabblock \,\subset\, {\cal N}(\stabblock) \,\subset\, {\cal N}(\stabH)$
and the quotient
\be
{\cal L} \definedby \faktor{\Pauli_{\nn}}{\normH}
\ee
Since ${\cal N}(\stabblock) \subset \normH,$ each coset in ${\cal L}$ is a union of cosets
of ${\cal N}(\stabblock)$: measuring the stabilizers of $\stabH$ partitions the Pauli group
into coarser error classes than measuring all of $\stabblock,$ as expected for a
compressive encoding.
\subsection{Main Result}\label{main_result}
The form of the generators of $\normH$ elucidates the structure of the quotient group ${\cal L}.$
The column generators have weight
$\wt{{\epi{\ell}\otimes\eta_a}} = \wt{\epi{\ell}}\cdot\wt{\eta_a} \geq  \qd,$ and we show
below (Corollary \ref{distance_cor}) that no element of
${\cal N}(\stabH) \setminus \stabblock$ has weight less than $\qd$: the distance of the
hybrid scheme is exactly the distance of $\qcode.$
Thus, we immediately recover the detectability condition of the quantum code $\qcode,$
namely, the number of errors in a logical qubit must be strictly less than $\qd$
to anti-commute and be detectable. For this reason, our construction
is not a quantum code, {\it per se}, but more aptly described
as a hybrid classical-quantum coding scheme.

Logical operations require no modification. The construction alters
syndrome extraction only: the encoded states remain stabilized by the full
product group $\stabblock,$ and the logical operators of the register are
those of the constituent code applied blockwise, $\epi{\ell} \otimes \eta$
for $\eta \in \normS \setminus \stabgroup,$ or
$\id_\numL \otimes {\cal L}_{\qcode}$ with ${\cal L}_{\qcode}$ the logical
algebra of $\qcode.$ Logical gates are therefore implemented exactly as in
the canonical scheme. The compression acts on the checks, with
$\id_\numL \kron \hq$ replaced by $\hl \kron \hq,$ and never on the logical
operators. The measured subgroup $\stabH$ itself should not be read as the
stabilizer group of a new code; the distinction is taken up in the remark
following Corollary \ref{distance_cor}.

Let $\ep_{\ast \ell}$ denote the $\ell$-th column of $\ep$ (the error pattern 
on the $\ell$-th logical qubit) and  $\colwt{\ep}$  the number of non-zero columns of $\ep.$ 
A Pauli operator that negates either normalizer condition (\ref{ep_conditions})
anti-commutes with $\stabH$ and is detectable. Roughly, if $\colwt{\ep} < \ld,$
the non-zero rows of $\hq \ep$ are too light to be codewords of ${\cal C},$
and a non-stabilizer column with $\wt{\ep_{\ast \ell}} < \qd$ is detectable
within its block. These observations are made precise by the lemmas below.
The main result is simply stated:

\vspace{\baselineskip}
 {\it A lookup table decoder exists 
for an encoding with $\hl \otimes \hq$ if the number of logical qubits with
errors (of each error type) is not greater than $\ltt$ and the number of errors in each 
logical qubit is not greater than $\qtt.$}
\vspace{\baselineskip}

Our claim is that all error patterns from the set
\be
\begin{split}
\mathbb{E} = \big\{ \ep \in \left\{\epX,\epZ \right\} \mid \; &\wt{\ep_{\ast \ell}} \leq \qtt \;\; \forall \ell , \\
&\colwt{\ep} \leq \ltt \big\}
\end{split}
\label{error_class}
\ee
are correctable. The first condition $\wt{\ep_{\ast \ell}} \leq \qtt$ is
familiar from QEC theory. Coding via $\hl \otimes \hq$ adds
the second condition enforcing a constraint  {\it across} the block of logical qubits.
The proof rests on two elementary lemmas. The first expresses the
detectability afforded by the distance of the quantum code, and the second
collects standard properties of the outer code.

\begin{lemma}\label{inner_lemma}
Let $v \in {\mathbb F}_2^\numQ$ be non-zero with $\wt{v} < \qd.$ If
$\hq v = 0,$ then $v$ is the binary representation of a stabilizer element of
the given error type ($X$ or $Z$); in particular, if $v$ does not represent a
stabilizer element, then $\hq v \neq 0.$
\end{lemma}

\begin{proof}
Suppose $\hq v = 0.$ Then the corresponding Pauli operator of the given error
type commutes with every stabilizer generator of the opposite type; since
same-type Pauli operators commute in a CSS code, it lies in $\normS.$ The
minimum weight of the elements of $\normS \setminus \stabgroup$ is the code
distance $\qd,$ and $\wt{v} < \qd,$ so it lies in $\stabgroup.$
\end{proof}

\begin{lemma}\label{outer_lemma}
Let $w, w^\prime \in {\mathbb F}_2^\numL.$
(i) If $w \neq 0$ and $\wt{w} < \ld,$ then $w \hl^\tp \neq 0.$
(ii) If $\wt{w}, \wt{w^\prime} \leq \ltt$ and $w \hl^\tp = w^\prime \hl^\tp,$
then $w = w^\prime$; that is, every vector of weight at most $\ltt$ is the
unique coset leader of its coset.
\end{lemma}

\begin{proof}
(i) If $w \hl^\tp = 0,$ then $w \in \lcode$; every non-zero codeword of
$\lcode$ has weight at least $\ld,$ a contradiction. (ii) The sum
$w \oplus w^\prime$ lies in $\lcode$ and has weight at most
$2 \ltt < \ld,$ hence vanishes by the same argument.
\end{proof}

The main result now follows.

\begin{theorem}\label{main_theorem}
Let $\ep, \ep^\prime \in {\mathbb E}$ have equal product syndromes,
$\hq \ep \hl^\tp = \hq \ep^\prime \hl^\tp.$ Then every column of
$\Delta \definedby \ep \oplus \ep^\prime$ is the binary representation of a
stabilizer element of $\qcode$; that is, the Pauli operators corresponding to
$\ep$ and $\ep^\prime$ differ by an element of the product stabilizer group
$\stabblock$ and are corrected by a common recovery operator. Consequently, a
lookup table decoder exists for the error class ${\mathbb E}.$
\end{theorem}

\begin{proof}
The column and column-weight bounds defining ${\mathbb E}$ give
\bea
\wt{\Delta_{\ast \ell}} \leq \wt{\ep_{\ast \ell}} + \wt{\ep^\prime_{\ast \ell}}
\leq 2\qtt < \qd \quad \forall \ell \nonumber \\
\colwt{\Delta} \leq \colwt{\ep} + \colwt{\ep^\prime} \leq 2\ltt < \ld  \nonumber
\eea
Consider the matrix $\hq \Delta.$ By assumption
$\hq \Delta \hl^\tp = {\mathbf 0},$ so every row of $\hq \Delta$ lies in the
kernel of $\hl,$ that is, every row is a codeword of $\lcode.$ The support of each row of
$\hq \Delta$ is contained in the set of non-zero columns of $\Delta,$ hence
every row has weight at most $\colwt{\Delta} < \ld$ and must vanish by
Lemma \ref{outer_lemma}(i). Therefore $\hq \Delta = {\mathbf 0}.$
Now each column satisfies $\hq \Delta_{\ast \ell} = 0$ with
$\wt{\Delta_{\ast \ell}} < \qd,$ so by Lemma \ref{inner_lemma} each
$\Delta_{\ast \ell}$ is the binary representation of a stabilizer element of
$\stabgroup$ (possibly the identity). The corresponding Pauli operator on the
block is therefore an element of $\stabblock,$ and the Pauli operators of
$\ep$ and $\ep^\prime$ lie in the same coset of $\stabblock.$ Any recovery
operator correcting $\ep$ thus also corrects $\ep^\prime.$ A lookup table
decoder is obtained by tabulating, for each product syndrome attained on
${\mathbb E},$ a recovery operator for any error pattern in ${\mathbb E}$
producing that syndrome.
\end{proof}

For non-degenerate codes, the product syndrome map is injective outright.

\begin{corollary}\label{nondegenerate_cor}
If no non-identity stabilizer element of $\qcode$ has weight at most
$2\qtt$ (as for the non-degenerate Steane code), then $\Delta = {\mathbf 0}$
in Theorem \ref{main_theorem} and the product syndrome map
$\ep \mapsto \hq \ep \hl^\tp$ is injective on ${\mathbb E}.$
\end{corollary}

The lemmas also settle the distance claim made above.

\begin{corollary}\label{distance_cor}
Assume $\ld \geq \qd.$ The minimum weight of the elements of
${\cal N}(\stabH) \setminus \stabblock$ is $\qd.$
\end{corollary}

\begin{proof}
Let $N \in {\cal N}(\stabH)$ with $\wt{N} < \qd,$ and let $\epN$ denote the rectangular
representation of either error-type component of $N,$ so that $\wt{\epN} < \qd$ and
$\hq \epN \hl^\tp = {\mathbf 0}.$ Every column of $\epN$ has weight less than $\qd,$ and
$\colwt{\epN} \leq \wt{\epN} < \qd \leq \ld.$ The rows of $\hq \epN$ are codewords of
$\lcode$ supported on the non-zero columns of $\epN,$ hence of weight less than $\ld,$
and must vanish by Lemma \ref{outer_lemma}(i): $\hq \epN = {\mathbf 0}.$ By
Lemma \ref{inner_lemma}, every column of $\epN$ is then the binary representation of a
stabilizer element, so each component of $N,$ and therefore $N$ itself, lies in
$\stabblock.$ Conversely, for $\eta \in \normS \setminus \stabgroup$ of minimal weight
$\qd,$ the column generator $\epi{1} \otimes \eta$ has weight $\qd$ and lies in
${\cal N}(\stabH) \setminus \stabblock$ since its first column is not a stabilizer
element.
\end{proof}

It is instructive to separate the two types of generators of $\normH$
identified in Sec.~\ref{subgroups}. The column generators
$\epi{\ell} \otimes \eta_a$ commute with every stabilizer of the product
code and include the logical operators of the register; their minimal
non-stabilizer weight is $\qd,$ and they set the distance of the scheme.
The row generators $g_m \otimes E_q$ are different in kind. They
anti-commute with elements of $\stabblock$ and are therefore detectable
errors of the product code, invisible only to the measured subgroup
$\stabH.$ Every non-trivial product of row generators has its non-zero rows
in $\lcode,$ hence at least $\ld$ non-zero columns. Since any two patterns
of ${\mathbb E}$ differ in at most $2\ltt < \ld$ columns, no such element
can connect two patterns of the class. This is the exclusion effected by
Lemma \ref{outer_lemma}(i) in the proof of Theorem \ref{main_theorem}, and
it is the operational role of the distance of the outer code. Arbitrary
elements of $\normH,$ products of both types, are excluded by the two
mechanisms acting jointly. The distance $\qd$ is a worst case, attained by a
logical fault confined to a single block. In the best case, the class
${\mathbb E}$ contains clustered patterns of weight $\ltt \, \qtt,$ the
product of the two correction radii, so the scheme corrects far beyond its
distance in the manner of classical tensor product codes \cite{wolf1965codes}.

The proof of Theorem \ref{main_theorem} also yields a constructive,
two-stage decoding procedure that factorizes the lookup table for the pair
$\left( \lcode, \qcode \right)$ into standard tables for the constituent codes.
Since the support of each row of $\syndQ = \hq \ep$ is contained in the set of
non-zero columns of $\ep,$ each row of $\syndQ$ has weight at most
$\colwt{\ep} \leq \ltt$ and, by Lemma \ref{outer_lemma}(ii), is the unique
coset leader associated with the corresponding row of the measured product
syndrome $\syndCQ = \syndQ \hl^\tp.$ Decoding each row of $\syndCQ$ with the classical
decoder for $\lcode$ thus recovers $\syndQ$ exactly. The $\ell$-th column of
$\syndQ$ is then the ordinary $\qcode$-syndrome of the error pattern
$\ep_{\ast \ell}$ with $\wt{\ep_{\ast \ell}} \leq \qtt,$ which the decoder of
the constituent quantum code corrects (up to a stabilizer for degenerate
codes).

\section{Product Source Coding}\label{source_coding}
The basic construction is now cast in the framework of Shannon's source coding theorem. 
We view each data qubit as a memoryless random information source emitting 
symbols from the discrete alphabet of Pauli operators $\left\{ I, X, Y, Z \right\}.$ While not strictly
necessary, we model Pauli errors with independent Bernoulli random variables
$\left\{ {\cal X} , {\cal Y}, {\cal Z} \right\}$ with probabilities 
$p_x = P({\cal X}=X), p_y = P({\cal Y}=Y), p_z = P({\cal Z}= Z),$
respectively.  When  $ p_x = p_y = p_z,$ our error model is equivalent 
to the familiar depolarizing channel by source-channel coding duality, as noted in the Introduction. 
All quantum error correcting codes are compressive when viewed as a mapping from error patterns
to binary syndrome measurements: for each error type, the Steane 
$\QEC{7}{1}{3}$ code compresses a 7 dimensional error vector into 3 syndrome bits, 
and $\QEC{17}{1}{5}$ topological color codes compress a 17 dimensional error vector into 
8 syndrome bits.  Here, we stress the compressive properties 
of quantum and classical error correcting codes and refer to an encoding with 
$\hl \otimes \hq$ as {\it quantum error source coding} or, more briefly,
{\it product source coding}.

In his seminal work \cite{shannon1948mathematical}, Shannon defined a precise measure of the 
information content of a random source $\cal{X}$ in terms of a logarithmic function  $H_2(\cal{X}),$
commonly referred to as the {\it Shannon entropy}.  Informally,  Shannon's source
coding theorem states that $N$  independent, identically distributed (i.i.d.)
random variables  $\cal{X}$  can be compressed into 
$N H_2(\cal{X})$ bits with negligible probability of information loss. 
It is often remarked that quantum error correction is analogous to a heat 
engine transferring entropy from data qubits, hot with noise, to cold syndrome qubits. 
By the use of a classical error correcting code across a block of logical qubits, 
our construction makes this notion concrete and shows that the overhead 
needed for syndrome extraction for the entire computer 
is ultimately limited by the Shannon entropy 
of the error source, which is typically much less than 1.  

Any binary, linear code can be used in the construction.
For example, we may choose $\lcode$ from the Bose-Chaudhuri-Hocquenghem (BCH) 
family of codes \cite{peterson1972error} and $\qcode$ as the Steane code $\left( \qtt = 1 \right).$ 
While not achieving Shannon's source compression limit, the BCH family has the 
asymptotic code parameters $\left[ \numL, \numL - \ltt \lceil \log_2(\numL+1) \rceil, 2\ltt+1 \right].$
Using a $\ltt$-error correcting BCH code with large enough $\numL,$ 
our construction requires only ${\cal O} (\log_2 (\numL+1 ))$ extra
qubits for syndrome extraction. In the canonical approach to QEC, each logical 
qubit operates independently and  $\numL$ Steane logical qubits functioning as a block code 
corresponds to an $\numL$-error correcting code in our framework.  Here, we limit the 
number of logical qubits with errors to $\ltt \ll \numL,$  but achieve an exponential reduction in the 
number of syndrome qubits needed to perform quantum error correction, asymptotically.

\begin{figure}[t!]
\begin{center}
\includegraphics[width=0.45\textwidth]{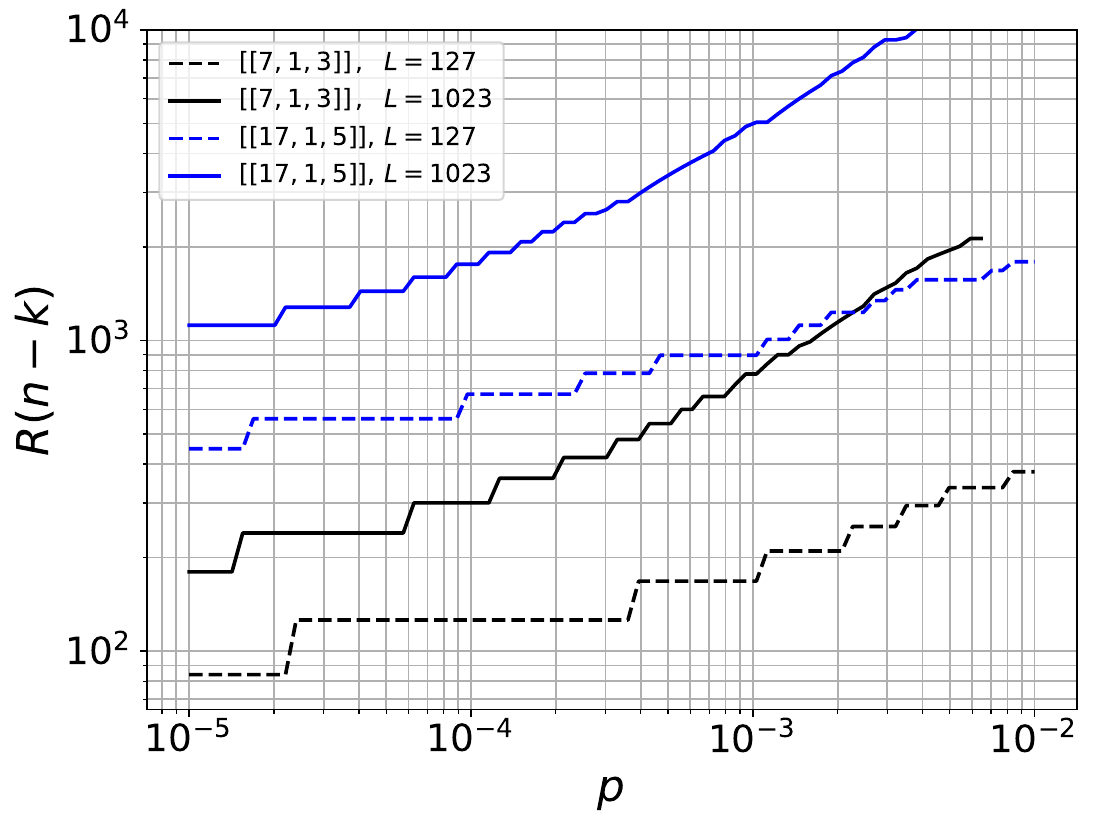}
\caption{Overhead plots for product source coding with 
$\lparams$ BCH codes and the $\steanecode$ Steane and $\colorcode$
color codes.  The number of syndrome qubits  $R \cdot (\numStab)$, where
$R=\numL - M$  and $(\numStab) = \card{\stabgroup},$  is plotted versus the physical error rate $p$. }
\label{overhead_fig}
\end{center}
\end{figure}

Syndrome qubit overhead from constructions formed by $\lparams$ BCH codes and the 
$\steanecode$ Steane and $\colorcode$ color code are shown in Fig. \ref{overhead_fig}.  The BCH
codes were chosen from  families with $\numL = 127$ and $\numL = 1023$ with sufficient 
distance such that the
failure probability is close to its minimum (see Sec. \ref{failure_prob}).  The
product code overhead  $R \cdot (\numStab)$ compares favorably against the
canonical approach where $\numL \cdot (\numStab)$ syndrome qubits are required.
For example, measuring 16 stabilizers of the color code from $\numL = 127$
logical qubits would require 2032 syndrome qubits in the canonical scheme, but
in the low noise regime, say $p=10^{-4}$, the BCH-color product 
code requires just 672. 
In general, compression by $\hl$ is characterized by the rate   $M / \numL $ of the classical code.  
The block length scaling of BCH codes 
 is evident as $\numL$ increases: at the error rate $p=10^{-4}$, 
 a computation with 1023 color code logical qubits requires
 an overhead of 1760 qubits, 
 less than a threefold increase over a system with 127 logical qubits.
 
\subsection{Failure Probability}\label{failure_prob}
The analysis of this section is at the {\it code-capacity} level: two-qubit
gates and measurements are assumed perfect, and errors afflict the data
qubits only. The lookup table decoder will fail to correct any error pattern not in ${\mathbb E},$
and thus the  protocol fails with probability
\begin{align}
P_F = \numL P( &\wt{\ep_{\ast \ell}} > \qtt ) + P(\colwt{\ep} >  \ltt )   \label{failure_prob_results}  \\
&- \numL  P( \wt{\ep_{\ast \ell}} > \qtt )P(\colwt{\ep} >  \ltt )  \nonumber
\end{align}
Assuming independent errors, 
the first term of $P_F$ is the probability that 
any logical qubit suffers errors exceeding $\qtt.$
For Bernoulli  sources with probability $p,$ we have 
\be
P( \wt{\ep_{\ast \ell}} > \qtt ) = 1 -  \sum_{\tau=0}^{\qtt} {\numQ\choose \tau} p^{\tau}(1-p)^{\numQ-\tau}
\label{P_L}
\ee
In the canonical scheme, $\numL P( \wt{\ep_{\ast \ell}} > \qtt )$ is an 
estimate of the failure rate, assuming perfect two-qubit gates and measurements.  
Table \ref{p_fail_code}  compares $P( \wt{\ep_{\ast \ell}} > \qtt )$ 
\footnote{Machine epsilon, defined 
as the smallest $\epsilon$ such that $1 + \epsilon > 1$, is ${\cal O}(10^{-16})$ for all numerical data}
for a Steane, color,  and $\golaycode$ Golay logical qubit. 

For a logical qubit with $\numQ$ data qubits 
and a Pauli error probability $p$, the probability of 
at least one error in a logical qubit is given by 
$
p_\ell =  1 - \left( 1 - p \right)^\numQ , 
$
and the second term in $P_F$ (\ref{failure_prob_results}) is binomial with probability $p_\ell$
\be 
P(\colwt{\ep} > \ltt ) = 1 -  \sum_{\tau=0}^{\ltt} {\numL\choose \tau} p_\ell^{\tau}(1-p_\ell)^{\numL-\tau}
\label{P_C}
\ee

Since error patterns exceeding the quantum correction radius $\qtt $
will cause a failure, the probabilities (scaled by $\numL$) in Table \ref{p_fail_code} 
serve as lower bounds on $P_F.$ Therefore, given an estimate of $p$, a good choice for the
classical code $\lcode$ is  one with sufficient distance such that  
\be
P(\colwt{\ep} > \ltt ) \sim {\cal O}\left( \numL  P( \wt{\ep_{\ast \ell}} > \qtt ) \right)
\label{P_good}
\ee
This methodology was followed to compute the overhead rates shown in Fig. \ref{overhead_fig}. 

\begin{figure}[t!]
\begin{center}
\includegraphics[width=0.45\textwidth]{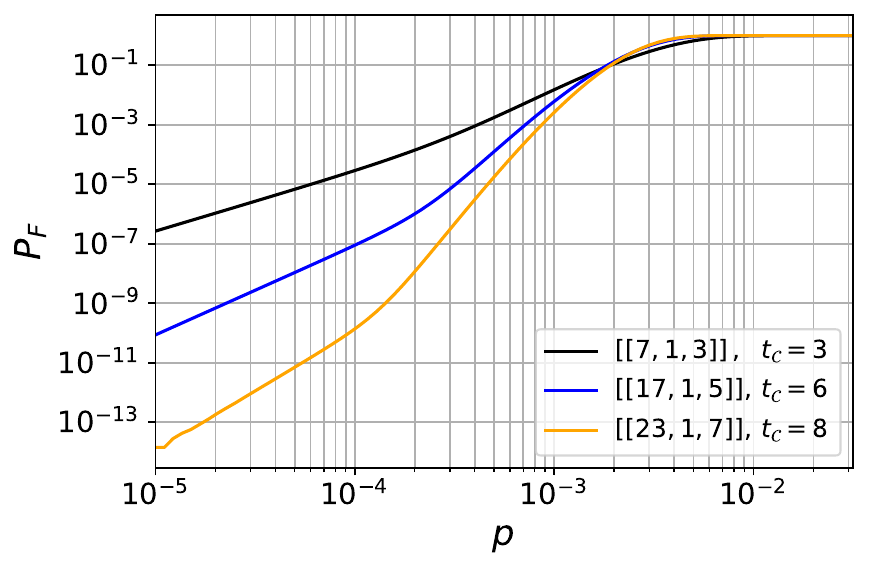}
\caption{Failure probabilities of  product source coding with $\numL = 127$ 
Steane, color and Golay logical qubits and a $\ltt$-BCH code plotted versus physical error 
rate $p$ assuming independent, Bernoulli errors.}
\label{P_F_t_fig}
\end{center}
\end{figure}

Failure probabilities  computed 
for  $\numL = 127$ with binomial probabilities (\ref{P_L}) and (\ref{P_C})  are plotted against
physical error rates  in Fig. \ref{P_F_t_fig}.  For each quantum code shown,  a single 
length-127 $\ltt$-BCH  code satisfying the criterion (\ref{P_good}) was chosen 
assuming $p=10^{-4}.$ From Fig. \ref{P_F_t_fig}, we observe that 127 color code
logical qubits achieve a  failure rate of ${\cal O}(10^{-7})$ at the cost of 672
syndrome qubits (from Fig. \ref{overhead_fig})  needed to correct both $X$ and $Z$-type errors 
occurring with probability $10^{-4}$.  As noted in the Introduction,  the inverse of the failure rate
is an estimate of feasible circuit depth.  Our failure probability accounts 
for {\it any} logical error, so here we have  $\Delta \sim 1/P_F,$ and observe 
that product code constructions with ${\cal O}(10^2)$ color or 
Golay logical qubits approach feasibility for running error-free circuits 
with depth $\Delta \ge {\cal O}(10^{9})$ at low physical error rates.


\begin{table}[t!]
\begin{ruledtabular}
\begin{center}
\begin{tabular}{cccc}
$\qcode$& $p=10^{-3}$ &  $p=10^{-4}$& $p=10^{-5}$  \\
\colrule
$\steanecode$  &  2e-05\,(3e-08) &  2e-07\,(3e-11)  &  2e-09\,(3e-14)   \\
$\colorcode$     &  7e-07\,(6e-12) &  7e-10\,(1e-16) &  7e-13\,(1e-16) \\
$\golaycode$    &  9e-09\,(2e-16) &  9e-13\,(1e-16) &  1e-16\,(1e-16) \\
\end{tabular}
\end{center}
\end{ruledtabular}
\caption{Comparison of probabilities of a high weight error pattern exceeding
the correction radius $P( \wt{\ep_{\ast \ell}} > \qtt ) $ and the 
distance $P( \wt{\ep_{\ast \ell}}  \geq \qd ) $ (in parentheses) for a
Steane, color and Golay  logical qubit with Pauli error rates  
$p=10^{-3}, 10^{-4},$ and $10^{-5}$ }
\label{p_fail_code}
\end{table}

\subsection{Noisy Syndrome Encoding}\label{noisy_encoding}
Faulty two-qubit gates and measurement errors
are likely to dominate in any quantum processor.  This remains the case in
current devices, where two-qubit gate and measurement error rates typically
exceed those of single qubit rotations or
random errors occurring while qubits are idle by an order of magnitude or
more.  Continuing within the source coding
framework, since two-qubit gates provide the syndrome encoding mechanism, 
we view two-qubit errors as {\it encoding} errors,  assumed to be 
generated by a Bernoulli source ${\cal E}$ with probability $p_e.$ To simplify 
matters, assume that ${\cal E}$ affects a (classical) bit-flip in the syndrome
measurement outcome and does not leave additional errors in the data
qubits.
This is the {\it phenomenological} noise model, here with check-weight-dependent
measurement error rates as quantified below.

The product code construction can be adapted to identify errors in the 
measured syndromes by using  a higher distance classical code and encoding 
fewer logical qubits.  To this end, recall that the generator matrix $\gg$ of a classical 
$\left[ n, k, d \right]$ code maps a {\it message} to a {\it codeword} 
by appending the message to its syndrome. 
For $\gg = \left[ \, \pl \, \, | \, \,\id_k \, \right]$ in systematic form, 
a message vector ${\mathbf m}$  is encoded as
\be
{\mathbf m} \gg = \left[ \, {\mathbf m}\pl \, \, | \, \, {\mathbf m} \, \right]
\label{classical_encoding}
\ee 
We exploit the defining properties of classical codes, namely
\be
\wt{{\mathbf m} \gg  \oplus {\mathbf m}^\prime\gg } \geq d
\label{classical_distance}
\ee
for  ${\mathbf m} \not = {\mathbf m}^\prime$
and an error corrupted codeword ${\mathbf m} \gg \oplus {\mathbf z}$
with $\wt{{\mathbf z}}\leq  \left \lfloor \frac{d-1}{2} \right \rfloor = t$ is 
closer to ${\mathbf m} \gg$ (in Hamming distance) than it is to 
any other codeword. 

Express the parity-check matrix of $\lcode$ in systematic form
$\left[ \, \id_{\lred} \, \, | \, \, \pl^\tp \,\right].$ For this construction, $\hl = \pl^\tp$
must be used (thereby reducing the size of the block of logical qubits to
$\lk$), so that error patterns are $\numQ \times \lk$ binary matrices and the
measured product syndrome is $\syndCQ = \syndQ \pl$ with syndrome matrix
$\syndQ = \hq \ep.$
Let $t_{src} < \ltt$ denote the maximum number of source errors, that is, the
maximum number of logical qubits with errors, and define the restricted error
class ${\mathbb E}_{src}$ of error patterns with at most $\qtt$ errors per
logical qubit and at most $t_{src}$ non-zero columns.
The key is to view the rows of $\syndQ$ as ``hyper-messages''  encoded in a
manner analogous to the classical case (\ref{classical_encoding}), but with the important
distinction that in the quantum setting the messages are
never transmitted,  only their syndromes are measured and used by the decoder.
The next theorem makes the protection afforded by this encoding precise.

\begin{theorem}\label{noisy_theorem}
Let $\ep, \ep^\prime \in {\mathbb E}_{src}$ have distinct syndrome matrices,
$\syndQ = \hq \ep \neq \hq \ep^\prime = \syndQ^\prime.$ Then the
measured product syndromes satisfy
\be
\wt{\syndCQ  \, \oplus \,  \syndCQ^\prime} \geq \ld - 2\,t_{src} .
\label{product_synd_distance}
\ee
Consequently, an erroneous measurement $\syndCQ \oplus \epT,$ for a
measurement error pattern $\epT$ with
\be
\wt{\epT} \leq  \left \lfloor \frac{ \ld - 2 t_{src} - 1}{2} \right\rfloor = \ltt - t_{src} ,
\label{meas_budget}
\ee
is strictly closer in Hamming distance to $\syndCQ$ than to any other product
syndrome arising from ${\mathbb E}_{src},$ and minimum distance decoding
recovers $\syndCQ$ uniquely.
\end{theorem}

\begin{proof}
Since $\syndQ \neq \syndQ^\prime,$ the two matrices differ in at least one
row; fix such a row index $i.$ The support of each row of $\hq \ep$ is
contained in the set of non-zero columns of $\ep,$ so
$\wt{\syndQ_{i\ast}} \leq \colwt{\ep} \leq t_{src}$ and, likewise,
$\wt{\syndQ^\prime_{i\ast}} \leq t_{src}.$ By (\ref{classical_encoding}), the vectors
$\left[ \, \syndQ_{i\ast}\pl \, \, | \, \, \syndQ_{i\ast} \, \right]$ and
$\left[ \, \syndQ^\prime_{i\ast}\pl \, \, | \, \, \syndQ^\prime_{i\ast} \, \right]$ are distinct
codewords of $\lcode,$ so (\ref{classical_distance}) and the component-wise
definition of Hamming distance give
\be
\wt{\syndQ_{i\ast} \pl \oplus \syndQ^\prime_{i\ast} \pl}
\geq \ld - \wt{\syndQ_{i\ast} \oplus \syndQ^\prime_{i\ast}}
\geq \ld - 2\,t_{src} .
\nonumber
\ee
The left-hand side is the distance between the $i$-th rows of $\syndCQ$
and $\syndCQ^\prime,$ and the full product syndromes inherit this separation
since $ \wt{\syndCQ  \, \oplus \,  \syndCQ^\prime} \geq
\wt{\syndCQ_{i\ast} \, \oplus \,  \syndCQ^\prime_{i \ast}},$ establishing
(\ref{product_synd_distance}). For the decoding claim, if
(\ref{meas_budget}) holds, then $\syndCQ \oplus \epT$ lies within distance
$\ltt - t_{src}$ of $\syndCQ,$ while its distance to any other product
syndrome is at least
$\left(\ld - 2 t_{src}\right) - \left(\ltt - t_{src}\right) > \ltt - t_{src}$
since $\ld \geq 2 \ltt + 1.$
\end{proof}

Note that the errors comprising $\epT$ may occur anywhere in the product
syndrome $\syndCQ.$ Error patterns in ${\mathbb E}_{src}$ with identical
syndrome matrices $\syndQ$ require no discrimination: by
Theorem \ref{main_theorem}, they differ by an element of the product
stabilizer group and are corrected by a common recovery operator.

Once the expected maximum number of source errors $t_{src}$
is determined, decoding for the error class 
\be
\overline{\mathbb{E}} = \mathbb{E} \cup \big\{   \epT  \, \big | \, \wt{\epT} \leq \ltt - t_{src} \big\}
\ee
is performed by nearest (in Hamming distance)
neighbor search on the keys (product syndromes) of the lookup table for $\mathbb{E}.$ 
This {\it minimum distance decoder} turns out to be the maximum-likelihood decoder
for independent Bernoulli  error sources \cite{peterson1972error}. Moreover, there are a number of
classical data structures and algorithms that efficiently perform minimum
distance search in metric spaces by exploiting the triangle inequality, such as a BK-tree
\cite{burkhard1973some}, that can be employed for minimum distance decoding
with a lookup table. The robustness afforded by Theorem \ref{noisy_theorem} is
a one-round guarantee against classical flips of the measured syndrome bits,
and is thus complementary to single-shot quantum error correction
\cite{bombin2015single,campbell2019theory}, where redundancy among the
stabilizer checks themselves protects against measurement noise. In our
construction, that redundancy is supplied explicitly by the distance of the
classical code $\lcode.$

Assuming i.i.d. two-qubit errors, the probability of an encoding
error in a syndrome qubit scales with the number of two-qubit 
gates coupling to it.   Thus the weight  (or {\it density}) of a stabilizer
 $h_\idxR \otimes s_\idxS,$ denoted $\delta_{\idxS \idxR} = 
 \wt{h_\idxR \otimes s_\idxS},$ is the relevant quantity
 in computing the failure probability due to encoding errors.
Since an even number of two-qubit errors in a syndrome qubit is self-correcting, the 
$(\idxS, \idxR)$-th syndrome measurement error probability
is given by the series 
\be
P_{\idxS\idxR} = \sum_{a \in O} {\delta_{\idxS \idxR} \choose a} p_e^a (1-p_e)^{\delta_{\idxS \idxR} - a}
\label{delta_prob}
\ee
where $O = \left\{ 1, 3,  \, \dots \right\}$ is the set of odd integers not greater than $\delta_{\idxS\idxR}$
and the probability of an uncorrectable measurement error  $P( \wt{\epT} > \ltt - t_{src})$
follows a Poisson binomial distribution with probabilities $\left\{ P_{\idxS \idxR}\right\}.$

\subsection{Error Detection and Localization}
Recall from Lemma \ref{inner_lemma} that a non-stabilizer column with
$\wt{\epArg{D}_{\ast \ell}} < \qd$ produces a non-zero $\qcode$-syndrome and
is detectable. With a syndrome 
decoder constructed from $\hl$ only, a localization method 
can be devised for the set of error patterns
\be
\mathbb{D} = \Big \{ \epArg{D} \in \big\{\epX,\epZ \big\}  \,\Big | \,\wt{\epArg{D}_{\ast \ell}} < \qd ,  
\colwt{\epArg{D}} \leq \ltt \Big \}
\ee
Since $\mathbb{D}$ contains detectable error patterns beyond the correction radius for $\qcode,$
there exist error patterns $\epArg{D} \ne \epArg{D}^\prime$ with identical product
syndromes $\syndCQ = \syndCQ^\prime,$ and the unambiguous
lookup table decoder cannot be directly applied here. However, detectability of
logical qubit errors is sufficient for localization. Let
$\mathbb{L} \subseteq \idxset{\numL}$ denote the set of indices $\ell$ for
which the column $\epArg{D}_{\ast \ell}$ is not the binary representation of
a stabilizer element---the logical qubits carrying non-trivial errors (for
degenerate codes, a column pattern lying in the stabilizer acts trivially on
its logical qubit and requires no correction).
Row-wise decoding of the measured product syndrome recovers $\mathbb{L}$
exactly.

\begin{theorem}\label{localization_theorem}
Let $\epArg{D} \in \mathbb{D}$ and $\syndQ = \hq \epArg{D}.$ Decoding the
$\idxS$-th row of the measured product syndrome $\syndCQ = \syndQ \hl^\tp$
with the decoder for $\lcode$ recovers the row $\syndQ_{\idxS \ast}$ exactly,
and
\be
\mathbb{L} = \bigcup_{\idxS} \mathrm{supp} \left( \syndQ_{\idxS \ast} \right)
\ee
where $\mathrm{supp}(v)$ denotes the set of non-zero indices of $v.$
\end{theorem}

\begin{proof}
The support of each row $\syndQ_{\idxS \ast}$ is contained in the set of non-zero columns
of $\epArg{D},$ so $\wt{\syndQ_{\idxS \ast}} \leq \colwt{\epArg{D}} \leq \ltt,$ and by
Lemma \ref{outer_lemma}(ii) each $\syndQ_{\idxS \ast}$ is the unique coset leader
consistent with the corresponding row of $\syndCQ$; row-wise decoding
therefore recovers each $\syndQ_{\idxS \ast}$ exactly. For the support identity,
let $\ell \in \mathbb{L}.$ Then $\epArg{D}_{\ast \ell}$ does not represent a
stabilizer element and $\wt{\epArg{D}_{\ast \ell}} < \qd,$ so
Lemma \ref{inner_lemma} gives $\hq \epArg{D}_{\ast \ell} \neq 0$: column
$\ell$ of $\syndQ$ is non-zero and $\ell \in \mathrm{supp}(\syndQ_{\idxS \ast})$ for some
$\idxS.$ Conversely, if $\ell \in \mathrm{supp}(\syndQ_{\idxS \ast})$ for some $\idxS,$
then $\hq \epArg{D}_{\ast \ell} \neq 0,$ so the error on the $\ell$-th
logical qubit anti-commutes with the stabilizer group; in particular
$\epArg{D}_{\ast \ell}$ does not represent a stabilizer element and
$\ell \in \mathbb{L}.$
\end{proof}

In contrast to error correction on ${\mathbb E},$ which constrains each
logical qubit to at most $\qtt$ errors, exact localization tolerates any
detectable error pattern on the affected logical qubits.

For example, consider a product code formed by a block of 15 Steane logical
qubits and a $3$-error correcting BCH code $\lcode \sim \left[ 15,5,7 \right].$
Under the action of $\hq,$ the weight-5 error pattern 
\benonum
\includegraphics[width=0.25\textwidth]{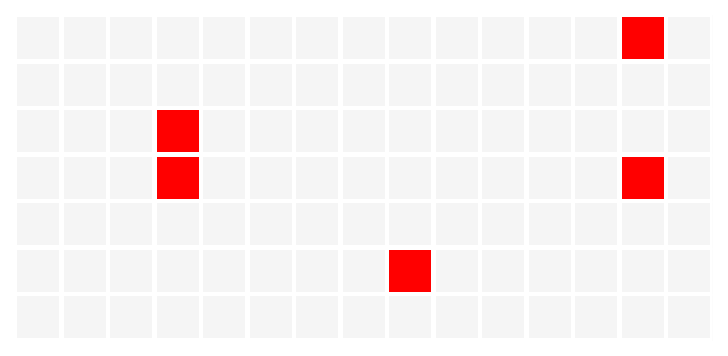}
\eenonum
with $\mathbb{L}  = \left\{ 4, 9, 14\right\}$ transforms to the $\qcode$-syndrome
\benonum
\includegraphics[width=0.25\textwidth]{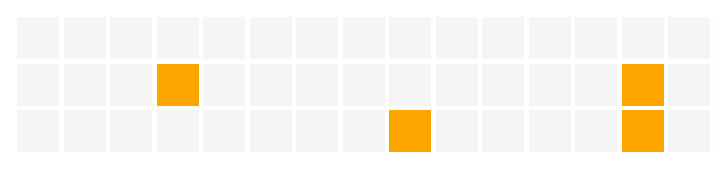}
\eenonum

Again, we interpret the rows of $\hq \epArg{D}$ as messages compressed by
$\hl.$ By Theorem \ref{localization_theorem}, a decoder designed for $\hl$
(such as a lookup table) applied to each row of the measured product syndrome
unambiguously recovers the full index set $\mathbb{L}$ of logical qubits with
errors.

Returning to the example, decoding the rows of the product syndrome
 $\hq \epArg{D}\hl^\tp$ illustrated as
\benonum
\includegraphics[width=0.175\textwidth]{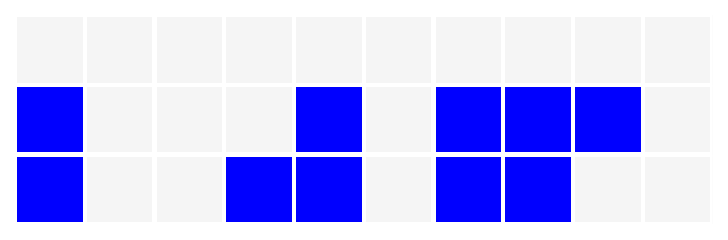}
\eenonum
yields the row index sets $\left\{ \{ \, \} , \, \{4, 14\} , \, \{9, 14\}  \right\},$ from which we 
conclude that logical qubits $\left\{ 4, 9, 14\right\}$ contain errors. 
To complete the error  correction cycle using the localization 
results, syndrome extraction using $\hq$ from each of
the logical qubits in $\mathbb{L}$ determines which data qubits contain errors.

We can exploit source-channel coding duality to identify both data qubit errors 
and errors in the syndrome measurements by encoding $M$ logical qubits and 
using a channel decoder designed for $\lcode \sim \lparams.$
Simply append a length-$M$ zero message to a row of the 
measured product syndrome  $ \syndCQ_{i\ast}  \oplus \epT_i$ to 
form the codeword $\left[  \, \syndCQ_{i\ast}  \oplus \epT_i \, | \, {\bf 0} \, \right]$ and use 
a classical channel decoder.  For $\lcode$ in the BCH family, an algebraic
decoder such as the  Berlekamp-Massey algorithm \cite{berlekamp1968algebraic}
will correctly identify errors in the logical qubits by interpreting
the zero message as erroneous and return the length-$M$ reconstruction of the
``message'' corresponding to  $ \syndCQ_{i\ast}.$ The support of the 
reconstructed ``message" will be contained in $\mathbb{L}.$
As a channel decoder, protection against
errors extends to the syndrome if 
$\wt{\epT_i} +   \card{ \mathbb{L}}   \leq \ltt.$ Sequentially decoding
all the rows of the product syndrome will unambiguously recover $\mathbb{L},$
provided $\epArg{D} \in \mathbb{D}.$

\begin{figure}[t!]
\begin{center}
\includegraphics[width=0.45\textwidth]{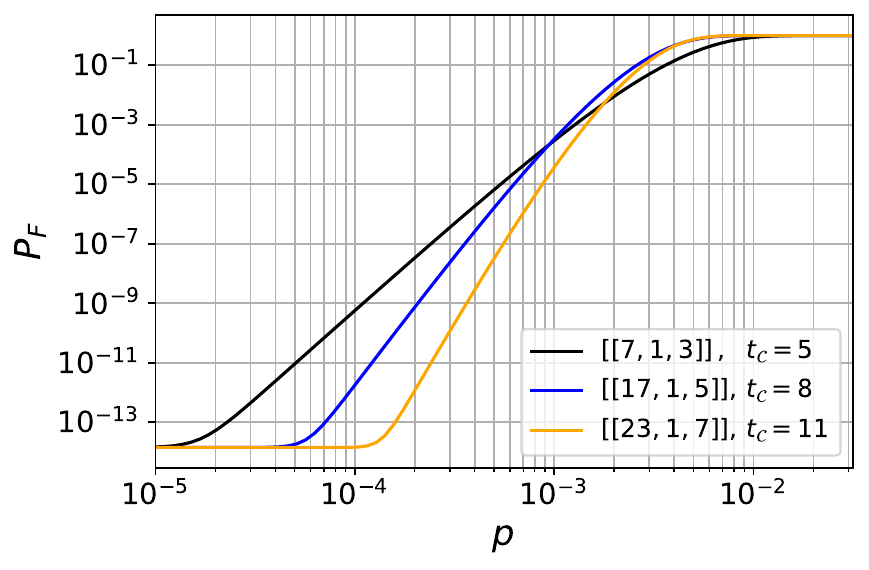}
\caption{Localization failure probabilities of  product source coding with $\numL = 127$ 
Steane, color and Golay logical qubits and a $\ltt$-BCH code plotted versus physical error 
rate $p$ assuming independent, Bernoulli errors.}
\label{P_F_d}
\end{center}
\end{figure}

Logical qubit localization holds advantages over error correction
since $ \mathbb{D} \supset \mathbb{E} $, and we may use the much lower 
probabilities $P(\wt{\epArg{D}_{\ast \ell}} \geq \qd)$ (see Table \ref{p_fail_code}) 
in estimating the failure probability $P_F$. 
Fig. \ref{P_F_d}  shows the probability of a localization error in $\numL = 127$
 Steane, color, and Golay logical qubits versus physical error rate. For each 
 quantum code, a single $\ltt$-BCH code satisfying 
$ P(\colwt{\ep} > \ltt ) \sim {\cal O}\left( \numL  P( \wt{\ep_{\ast \ell}} \geq \qd ) \right)$
was chosen assuming $p = 10^{-4}.$
\vspace{12pt}

\begin{figure*}[t!]
\begin{center}
\includegraphics[width=0.8\textwidth]{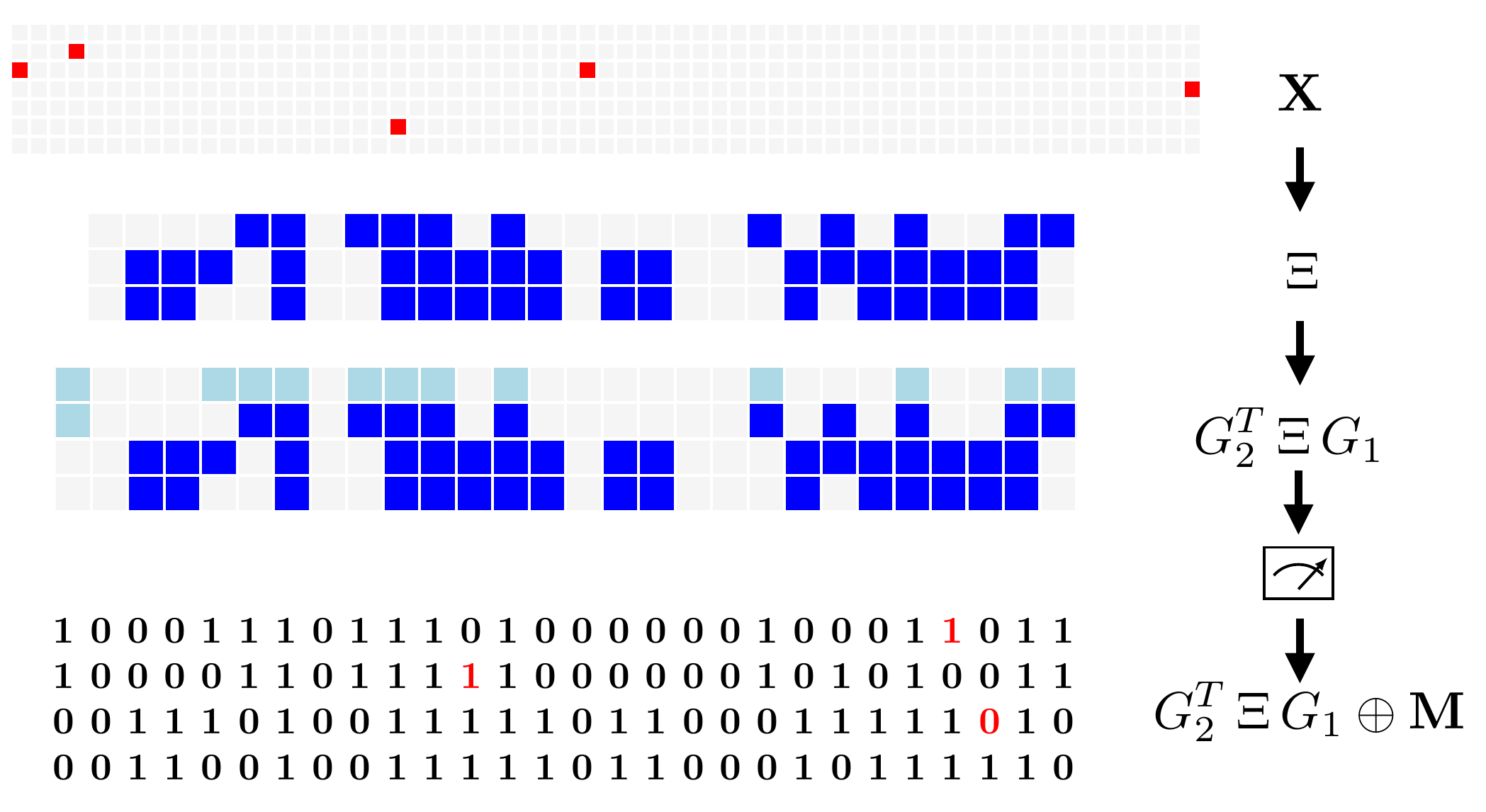}
\caption{Illustration of quantum error source and channel coding
with the Steane quantum code, a $\left[63,36,11\right]$ BCH classical source code, and a single 
parity-check product channel code. Identity operators in the rectangular
representation are white and non-identity are colored. From top: a weight-5 error 
pattern $\epX$ on 63 Steane logical qubits is compressed with a 
5-error correcting BCH code to a $3\times27$ source syndrome 
$\syndCQ= \hq^Z \epX \hl^\tp$ and coded for the measurement channel
with single-parity-check codes ${\mathbf G}_1$ and ${\mathbf G}_2.$ 
Bit-flips in the measurement outcomes due to $\epM$ are shown in red.}
\label{system_fig}
\end{center}
\end{figure*}

\section{Product Channel Coding}
A final construction completes a Shannon coding theory
for the quantum register by channel coding the compressed
product syndrome $\syndCQ.$ In this section, $\syndCQ$ plays the role of a
source-coded message---accordingly referred to as the {\it source
syndrome}---to be encoded by a classical error correcting code and sent
over a noisy channel, such as a quantum bus or measurement apparatus.
To this end, let  $\lcode_1 \sim \left[ n_1, \lred, d_1 \right]$ and 
$\lcode_2 \sim \left[n_2, (\numStab), d_2 \right]$ be classical codes and consider the 
binary matrix 
\be
 {\mathbf G}_1^\tp H_\lcode \otimes  {\mathbf G}_2^\tp H_\qcode
 \label{product_code}
\ee
with systematic generator matrices 
$\gg_1 = \left[ \, \pl_1 \, \, | \, \,\id_{\lred} \, \right] \, ,$ 
$\gg_2 = \left[ \, \pl_2 \, \, | \, \,\id_{(n-k)} \, \right]$  of 
$\lcode_1$ and $\lcode_2,$ respectively. 
The rows of  ${\mathbf G}_2^\tp H_\qcode$ are (modulo 2) linear combinations
of the rows of $H_\qcode$ and hence in $\stabgroup.$ As the Kronecker product of any
binary matrix with a quantum parity-check matrix is in the product stabilizer group,
we conclude that the rows of (\ref{product_code}) are in $\stabblock$ and therefore
suitable for quantum error syndrome extraction. For an error pattern $\ep,$ 
source-channel coding may be expressed as
\be
{\mathbf G}_2^\tp H_\qcode \ep H_\lcode^\tp {\mathbf G}_1 = {\mathbf G}_2^\tp \,\syndCQ \,{\mathbf G}_1
\ee
which may be arranged as the matrix  
\be
\left[
\begin{array}{cc}
 \pl^\tp_2 \, \syndCQ \,  \pl_1 & \pl^\tp_2 \, \syndCQ \, \\ 
 \syndCQ\, \pl_1 & \, \syndCQ
\end{array}
\right]
\label{product_mat}
\ee
where the submatrices 
$\left[\,\pl_2^\tp \,\syndCQ\,\, | \, \, \syndCQ \,\right]^\tp\in {\cal C}_2$ and 
$\left[\, \syndCQ \,\pl_1 \, |  \,\syndCQ\, \right] \in {\cal C}_1 .$ The remaining component 
$\pl_2^\tp \, \syndCQ \, \pl_1 $ is often referred to as 
  {\it check-on-checks} and is unique to product codes. The resulting product 
 code has parameters $\left[n_1 n_2 ,\lred \cdot (\numStab),d_1d_2 \right] .$ The matrix
 (\ref{product_mat}) is in the form of a (direct) product code as originally 
 proposed by Elias \cite{elias1954error} and recognized as the Kronecker product
 of the constituent codes 
 by Slepian \cite{slepian1960some}. The use of a classical error correcting code to 
 identify measurement errors in a single logical qubit was first proposed by Zalka 
 \cite{zalka1996threshold} using a single-parity-check code and more generally, 
 with an arbitrary classical error correcting code as described by Gottesman 
\cite{gottesman1997stabilizer} and attributed to unpublished work of Evslin, 
Kakade and Preskill therein.   In our construction, by virtue of the Kronecker
product, this procedure is extended from a single logical qubit to a block of logical qubits. 

The full source-channel product coding construction is illustrated in Fig. \ref{system_fig}. 
Analogous to the classical case, source and channel coding are depicted separately,
though these operations happen concurrently in the quantum setting.  Under the action of 
$\hl \otimes \hq^Z,$ the error pattern $\epX$ is compressed to source syndrome $\syndCQ$ (blue).    
Single-parity-check codes $\lcode_i$ with $\pl_i = \left[1, 1, \, \dots \, 1\right]^\tp$
compute parity checks across each row and column of $\syndCQ$ (light blue).  
Product channel coding adds the check-on-checks bit in the top left corner (light blue) 
and the source syndrome itself is sent through the measurement channel. 
Measurements corrupted by an error pattern $\epM$ flip bits in the observed binary outcomes (red). 

Decoding the channel code may be performed with a classical decoder to recover the 
noise corrupted source syndrome $\syndCQ,$ which may be queried against a lookup
table constructed from the pair $\left(\lcode,\qcode\right).$ Shannon's second theorem
\cite{shannon1948mathematical} characterizes the capacity 
of a discrete, memoryless channel as the limiting rate at which a message 
can be sent reliably through a noisy channel in terms of the maximum mutual 
information between the source (compressed error sources $\syndCQ$) and 
channel outputs 
(channel syndrome measurements ${\mathbf G}_2^\tp \,\syndCQ \,{\mathbf G}_1 \oplus {\mathbf M}$).  
For a binary symmetric measurement channel parameterized by $p_m$, the channel 
capacity is given by $1 - H_2(p_m)$. For the linear product codes considered here, 
Shannon's noisy-channel coding theorem yields the bound 
$\lred \cdot (\numStab) / n_1n_2 < 1 - H_2(p_m)$.

\section{Discussion}\label{discussion}
\subsection{Fault-Tolerance}\label{fault_tolerance}
Our construction violates the first law of fault-tolerant quantum error correction (FTQEC):
\textit{never use a syndrome qubit more than once} \cite{shor1996fault,preskill1998reliable}.  
Adherence to this law prevents an error in the preparation (or reuse)  of a syndrome qubit from 
propagating to a high weight, undetectable error pattern in a 
logical qubit.  As  proposed by Shor \cite{shor1996fault} and Steane \cite{steane1999efficient},
syndrome extraction may be made fault-tolerant by preparing blocks of syndrome qubits 
in an entangled state, coupling to the entangled block, and performing a parity 
measurement on the entangled block to obtain the syndrome bit.  In this way, 
each data qubit interacts with a single syndrome
qubit, preventing  a cascade of errors from subsequent couplings.   At the cost of extra syndrome qubits, errors  
in the syndrome qubits only propagate to low weight errors in the logical qubits which 
may then be caught and corrected in future rounds of error correction.  Central to our scheme, 
however, is the coupling of \textit{multiple} logical qubits to the \textit{same} syndrome  
qubit potentially exacerbating the propagation of errors. 

\begin{figure}[t!]
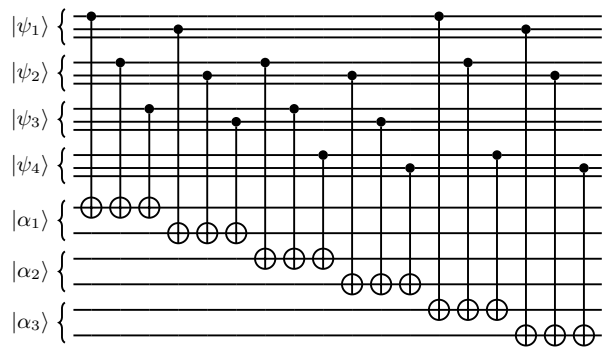

\resizebox{0.45\textwidth}{!}{\hammingbitflipzeroFT}
\caption{Fault-tolerant implementation of the stabilizer circuit of Fig. \ref{ZZI_circuit}.  
Ancilla blocks $|\alpha_i\rangle$ are prepared in Bell states. Fault-tolerant 
syndrome extraction is performed by coupling each data qubit in a logical qubit $(|\psi_j\rangle)$ 
to a separate ancilla qubit. Each Bell state ancilla block is coupled to multiple logical 
qubits, but a single error in an ancilla qubit propagates to a weight-1 error in each 
logical qubit to which it is coupled.}
\label{ZZI_circuit_FT}
\end{figure}

A full {\it circuit-level} analysis, in which every gate, state preparation, and
measurement may fail, is beyond the scope of this work, but we consider some key issues here.
Referring to the circuit in Fig. \ref{ZZI_circuit} and assuming that the
syndrome qubits are prepared in the $|0\rangle$ state, a $Y$\hspace{-2pt}-type error in the first 
syndrome qubit will propagate to a weight-2 $Z$-type error in the logical qubits 
$|\psi_1\rangle, |\psi_2\rangle$ and $|\psi_3\rangle.$ However, fault-tolerance may 
be recovered in the product coding scheme by adapting Shor's method \cite{shor1996fault}
as illustrated in Fig. \ref{ZZI_circuit_FT}.  
This circuit first prepares Bell states $| \alpha_i\rangle = (|00\rangle + |11\rangle)/\sqrt{2}$ 
(not shown in Fig. \ref{ZZI_circuit_FT})
and couples each data qubit in a logical qubit to a different qubit in a Bell state 
as prescribed by the classical and quantum error correcting codes used in the 
construction (i.e. $\hl \,\otimes \,\hq).$   
For example, in Fig. \ref{ZZI_circuit_FT}, the first data qubits in  
logical qubits $|\psi_1\rangle, |\psi_2\rangle$ and $|\psi_3\rangle$ are coupled to the first qubit in 
$| \alpha_1\rangle,$ and the second data qubits in  
logical qubits $|\psi_1\rangle, |\psi_2\rangle$ and $|\psi_3\rangle$ are coupled to the second qubit in 
$| \alpha_1\rangle,$ completing the couplings specified by the first row of $\pl^\tp \otimes [ \, 1 \, 1 \, 0 \,].$
The remaining rows of  $\pl^\tp \otimes [ \, 1 \, 1 \, 0 \,]$ are implemented similarly. 
As before, each logical qubit interacts with multiple Bell states, but 
coupling the qubits in this fashion ensures that a single fault in a Bell state propagates to a single
error in the logical qubits to which it is coupled, and thus retains the fault-tolerance property.  
To be fully fault-tolerant, one must repeat the measurement a number of times 
until convergence as described in \cite{shor1996fault}. 

To account for Shor style fault-tolerant syndrome extraction, the qubit overhead 
rates $R \cdot (\numStab)$ in Fig. \ref{overhead_fig} are scaled
by the weights of the stabilizers, which for the Steane code is 4 (see Appendix \ref{appsteane}),
yielding an overhead rate of  $4 R \cdot (\numStab)$.   The color code has 7 weight-4 stabilizers
and a weight-8 stabilizer for each error type (see Appendix \ref{appcolor}), and an overhead rate of
$4 R \cdot (\numStab-2) + 16 R$ when syndrome extraction is performed fault-tolerantly. 

\begin{figure*}[t!]
\includegraphics[width=0.65\textwidth]{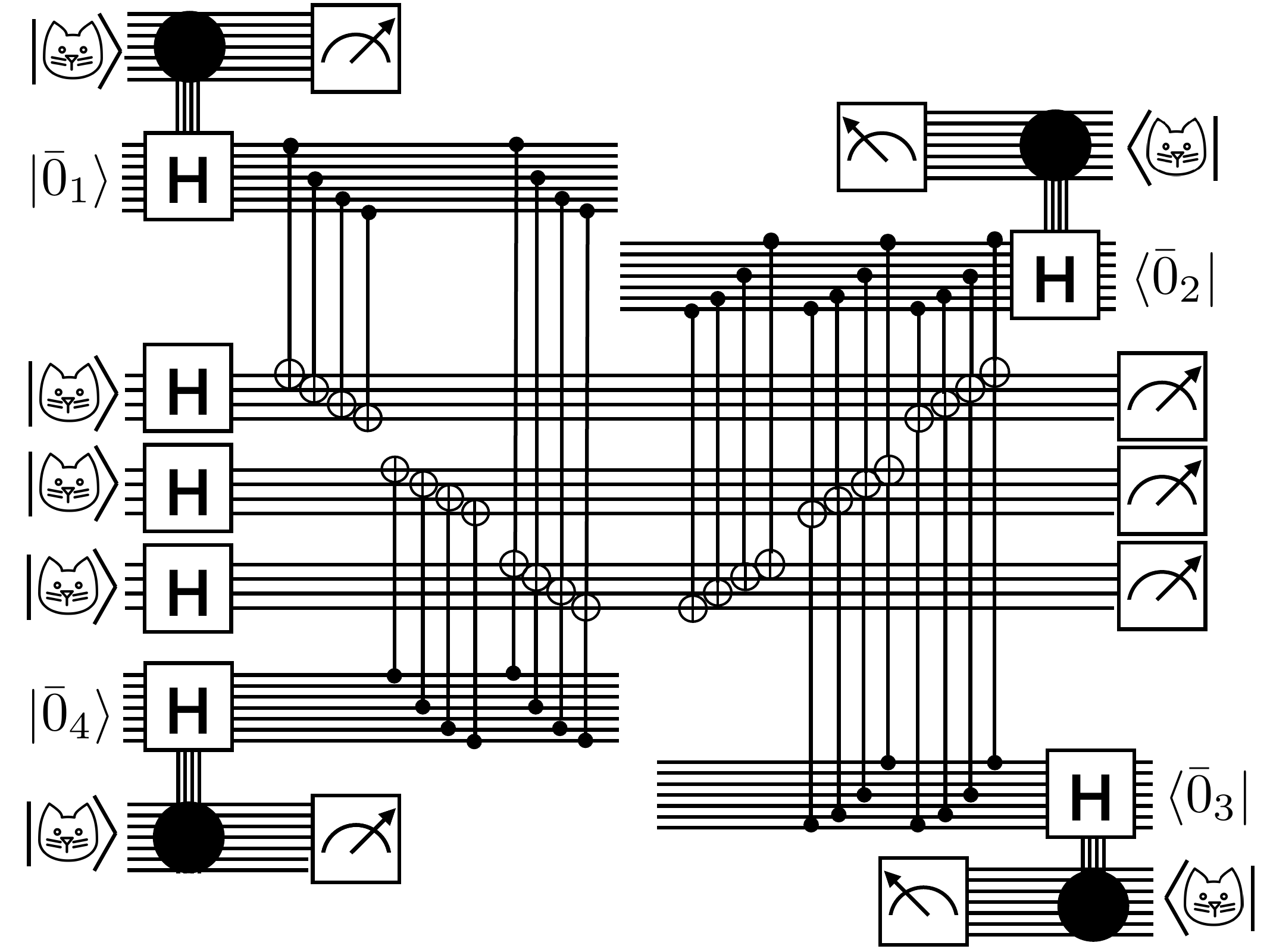}
\caption{Fault-tolerant encoded magic state factory. Cat states and Steane zero codewords 
are prepared and verified against weight-2 errors by ancilla factories (not shown). Cat
states are used to prepare a magic state $\pm | H \rangle$ by fault-tolerantly measuring 
a transversal Hadamard. CNOT gates with  common targets are overlaid for clarity.
Error detection and localization is performed fault-tolerantly
with additional cat states and Steane syndrome extraction, shown here by product
coding with  $\pl^\tp \otimes \left[ \, 1 \, 0 \, 0 \, 1 \, 0 \, 1\, 1\, \right]$.  Any magic states found with errors are discarded. 
The process iterates by cycling through the remaining Steane stabilizers
with a shortened classical parity-check matrix to couple the reduced number
of magic states.}
\label{t_state_factory}
\end{figure*}

\subsection{Ancillary Processes}
The product code construction applies to quantum processes that employ 
non-destructive measurements from data qubits to
ancillary qubits as a computational or post-selection primitive.  So-called {\it ancilla factories} are one
such application in which multiple-qubit entangled states (e.g. cat states, quantum codewords) 
are constructed and verified by measuring Pauli operators on the data qubits by coupling to ancillary qubits. 
Based on the measurement results of the ancillary qubits, the entangled state is accepted or discarded.
The basic principle proposed here---error extraction from blocks of entangled states
and collective inference---may be used to verify multiple entangled states
simultaneously. Moreover, for this type of post-selection task we may use the detection and localization
method as described previously, since any failure of a verification test leads to destruction of the state
undergoing verification---which qubit in the entangled state contains the error is not important.

Consider, for example, the verification of  $n$-qubit cat states of the form 
$\left(|00\cdots0\rangle + |11\cdots1\rangle\right)/\sqrt{2}.$ 
As illustrated by the 2-qubit cat (or Bell) states  in Fig. \ref{ZZI_circuit_FT}, 
these states facilitate fault-tolerant syndrome extraction.  More generally, $n$-qubit cat states 
transformed via transversal Hadamard gates create
even-parity states for use in fault-tolerant syndrome extraction for any quantum code. 
Cat states are verified  by  performing CNOT gates from the first and last qubits in the cat state to 
an ancilla qubit initialized as $|0\rangle.$  A non-zero measurement of the ancilla  
indicates that the cat state contains a weight-2 error  \cite{preskill1998reliable}, and 
is therefore not suitable for fault-tolerant operations and discarded. 
The $n$-qubit cat state verification circuit is represented by the  binary vector 
$V = \left[ \, 1 \, 0  \, \cdots \, 0 \,  1 \, \right],$ and an ancilla factory constructing 
multiple cat states may be verified by error localization and post-selection based 
on source coding with $\hl \, \otimes \, V.$ More generally, for example in the 
verification of quantum codeword encoding \cite{preskill1998reliable}, 
$V$ is a matrix composed of check operators arranged as rows. 

Magic states are key resource states in FTQEC as they complete the Clifford 
group of transformations to form a universal set of quantum logic operations \cite{PhysRevA.71.022316}.
The magic state $|A\rangle = \left(|0\rangle + e^{i\pi/4}|1\rangle \right)/\sqrt{2}$ can be used to
simulate a logical $T$ gate on encoded qubits,
and the set $\mathrm{Clifford} \,+\,T$ is known
to be universal for quantum computation.  Imperfect magic states can be iteratively 
improved by state distillation \cite{PhysRevA.71.022316},
however distillation techniques are not inherently fault-tolerant.  Alternatively, an {\it encoded} 
magic state can be constructed fault-tolerantly and used directly to simulate logical $T$ gates 
to achieve universality  \cite{10.5555/2011665.2011666}.  

A fault-tolerant encoded magic state factory can be built 
by combining the methods proposed in this work as illustrated in Fig. \ref{t_state_factory}. 
The parity-check matrix $\pl^\tp$ from the $\left[  7, \, 4 , \, 3 \right]$ 
Hamming code  (\ref{A_mat})  multiplexes error detection and localization from 4 encoded magic states,
chosen here to be $|H\rangle = \cos(\pi/8) | \bar{0}\rangle + \sin(\pi/8)| \bar{1}\rangle.$
The factory consumes 7-qubit cat states and Steane encoded zero states
$|\bar{0}_j\rangle, 1 \leq j \leq 4$. These states are the outputs of ancilla factories producing  cat states
and quantum codewords verified against weight-2 errors as described above.   Ancilla blocks 
prepared in cat states fault-tolerantly measure the Hadamard operator and project 
the Steane zero codewords $|\bar{0}_j\rangle$ onto 
a magic state $\pm | H\rangle$. The sign of the projected magic state is 
determined by a measurement of the cat state.  Verified 4-qubit cat states are 
consumed for fault-tolerant error detection and localization based on 
the stabilizers of the Steane code.  Transversal Hadamards  
first transform the 4-qubit cat states to an even parity states to carry out 
Shor style fault-tolerant syndrome extraction.  
The CNOT gates in Fig. \ref{t_state_factory} couple the errors to the ancilla blocks 
fault-tolerantly implementing the source code  $\pl^\tp \otimes \left[ \, 1 \, 0 \, 0 \, 1 \, 0 \, 1\, 1\, \right]$, 
corresponding to the Steane code stabilizer $Z_1 Z_4 Z_6 Z_7$.
Errors in the encoded magic states are localized by decoding the rows of $\syndCQ$ 
as described previously and those states are discarded.  The process iterates by
cycling through the remaining Steane code stabilizers, 
perhaps with a reduced number of magic states and discarding encoded magic states
found with errors.  To accommodate fewer encoded magic states, the classical 
code may be shortened by removing columns from $H_\lcode$. 

\subsection{Asymmetric and Correlated Error Models}
Our source coding constructions employed a single classical code for 
compression of $X$ and $Z$-type error sources.  With CSS codes, as  investigated here, one can choose 
separate codes for correcting $X$ and $Z$ errors to reflect asymmetry in the error rates. 
Dephasing is likely to dominate in qubits with a $Z$-type energy splitting, 
thus motivating interest in error correction protocols designed for asymmetric error models.
With an independent error model, the achievable compression rate of an 
encoding scheme with separate classical codes for each error type 
(e.g. $\lcode_{\cal X}$ and $\lcode_{\cal Z}$) is asymptotically limited by 
$H_2({\cal X}) + H_2({\cal Z})$ by Shannon's theorem.

A single data qubit may exhibit correlations between dephasing and bit-flip errors 
as observed in the paradigmatic depolarizing noise model \cite{delfosse2014decoding}.  
Source coding of correlated classical sources is characterized by the 
Slepian-Wolf theorem \cite{slepian1973noiseless}:  if a model of the correlations is 
known, Slepian and Wolf established the joint entropy $H_2({\cal X},{\cal Z}) \leq H_2({\cal X}) + H_2({\cal Z})$ 
as the achievable rate with separate coding for ${\cal X}$ and ${\cal Z}.$
Remarkably, knowledge of ${\cal X}$ is not needed to compress ${\cal Z}$ (and vice versa) but joint
decoding $\{ {\cal X}, {\cal Z} \}$ with the aid of a model can achieve the joint entropy rate.

Given our assertion that syndrome extraction performs 
{\it classical} data compression, by Slepian-Wolf, separate encodings for 
phase and bit-flip errors are possible (using, for example, the
implementations described in \cite{wyner1974recent, pradhan2003distributed}),
provided a  model of the correlated errors is known. Quantum noise spectroscopy
protocols have recently been extended to estimate multiple-axis noise correlations from experimental data
\cite{paz2019extending}, perhaps providing a path toward accurate correlated error models that
could be used in this setting. 

\subsection{Relation to Quantum LDPC Codes and Single-Shot Error Correction}\label{qldpc_relation}
It is instructive to situate the present construction within the rapidly
developing landscape of quantum LDPC codes and related protocols.  Modern
product constructions---hypergraph products \cite{tillich2013quantum},
balanced products \cite{breuckmann2021balanced}, and lifted products
\cite{panteleev2022asymptotically}---combine two classical codes into a
single quantum code whose stabilizer checks intertwine the constituents,
achieving constant rate and, remarkably, linear distance
\cite{panteleev2022asymptotically,leverrier2022tanner}.  High-rate quantum
LDPC memories have also been shown to substantially reduce qubit overhead in
near-term architectures \cite{bravyi2024high}.  Our construction is
complementary: the inner quantum code is left intact, and a classical code is
overlaid on a block of otherwise independent logical qubits.  The scheme
thereby preserves the transversal gates, decoders, and hardware layouts of
the constituent code while pooling syndrome extraction across the block, and
any improvement to either constituent---classical or quantum---transfers to
the product immediately.  The same distinction applies to syndrome noise:
single-shot error correction \cite{bombin2015single,campbell2019theory}
achieves robustness through inherent redundancy among the stabilizer checks
of a single code, whereas here the redundancy is imported explicitly through
the distance of the outer classical code, yielding the one-round guarantee of
Theorem \ref{noisy_theorem}.  Finally, the decoding advances developed for
quantum LDPC codes, notably belief-propagation with ordered-statistics
post-processing (BP-OSD) \cite{panteleev2021degenerate,roffe2020decoding},
apply directly to the rows of the measured product syndrome and offer a
practical alternative to lookup tables at scale.

\subsection{Bounds on Compression and Check Weight}\label{ldpc_bounds}
The check-weight cost stated in the Introduction can be made precise. For
the small code blocks considered in this work, $\numQ p \ll 1$ and the
block-error indicators form a sparse i.i.d. Bernoulli($p_\ell$) source, the
source the outer code compresses. For a $(w_c, w_r)$-regular LDPC outer
code, every product syndrome bit is a parity of $w_r$ source bits, and an
entropy argument dual to Gallager's \cite{gallager1963low,sason2003parity}
bounds the compression rate:
\be
\frac{\lred}{\numL} \geq \frac{H_2(p_\ell)}{H_2(\tilde{q})}
\quad \quad
\tilde{q} = \frac{1 - \left(1-2p_\ell\right)^{w_r}}{2}
\label{ldpc_bound}
\ee
The bound is essentially tight, since recovering an error pattern from its
syndrome is the same inference problem as decoding the outer code $\lcode$
over a binary symmetric channel \cite{gallager1963low,sason2003parity}, and
it carries the central conclusion of this section: compression within a
factor $(1+\epsilon)$ of the entropy limit requires check weight
$w_r = {\cal O}\left(p_\ell^{-1} \ln (1/\epsilon)\right),$ independent of
the block length $\numL.$ The algebraic codes employed in this work have
row weight $\Theta(\numL)$ instead.

For the $\qk = 1$ codes considered here the bound simplifies. Counting
both error types, the canonical scheme allocates
$(\numStab)/\numQ \approx 1$ syndrome qubits per data qubit and the product
construction $(\lred/\numL) \, (\numStab)/\numQ,$ so the boost over the
canonical scheme is $B = \numL/\lred,$ and (\ref{ldpc_bound}) caps it at
\be
B \leq \frac{H_2(\tilde{q}\,)}{H_2(p_\ell)} \leq \frac{1}{H_2(p_\ell)}
\approx \left[ \numQ p \, \log_2 \! \left( \frac{e}{\numQ p} \right) \right]^{-1}
\label{boost_bound}
\ee
the limit attained by a capacity-achieving classical code, with
$p_\ell \approx \numQ p$ in the closed form. At the design point of
Fig.~\ref{overhead_fig} the canonical scheme spends $16/17 = 0.94$ syndrome
qubits per data qubit; the BCH instantiation achieves $B = 3$ with product
checks of weight near 250, while (\ref{boost_bound}) permits $B = 5.7$ at
check weight 32 and $B = 55$ at capacity. The boost falls as $\numQ p$
grows, so pooling is most effective for small blocks at low error rates,
where the source is sparse. Approaching capacity requires unbounded check
density, and the guarantee there is a vanishing failure probability rather
than a correction radius. A product source coding scheme with
bounded-weight checks and belief-propagation decoding is developed in
forthcoming work.

\section{Conclusion}
In this work, we have proposed a versatile and efficient product code construction for syndrome
extraction from the encoded quantum register.  The construction connects Shannon's 
coding theorems and associated bounds to the overhead rates of QEC and other
quantum post-selection tasks. To demonstrate our method, we have 
investigated the BCH family of codes and lookup table
decoders for unambiguous (lossless) compression and error reconstruction. 
The size of the lookup table is  exponential in the number of allowable errors 
($\ltt$) and combinatorial in problem size ($\numL$), thus 
limiting the ultimate utility of a lookup table decoder. Nonetheless, a proof of concept design
reaching quantum advantage sized problems and circuits in the low noise
regime was presented.  Alternatively, the construction may be used to locate errors at the logical level.
This coarse-grained approach not only tolerates a higher number of 
errors per error detection cycle, but also allows for classical decoders
(paired with $\lcode$) to operate as syndrome decoders on the rows
of the measured product syndrome. Crucially for FTQEC, both decoding
paradigms---lookup tables and classical decoders for localization---are
robust to syndrome noise.

Improvements in decoding and  different classical encodings will likely accommodate higher error rates.   
In particular, the density of the classical code drives the 
tolerable two-qubit error rate by the dependence on the weight of the product stabilizers  
in the probability of an encoding error. Algebraic codes, such as the BCH family,  
are {\it high density} codes, comprised of high weight parity-check constraints. 
The construction with a classical {\it low-density parity-check} (LDPC) code
\cite{gallager1962low} would therefore reduce the number of two-qubit gates
needed for syndrome extraction; indeed, by the bound (\ref{ldpc_bound}),
compression near the entropy limit is compatible with check weights
independent of the block length.  A number of
deterministic and random LDPC constructions, as well as probabilistic belief-propagation decoders
\cite{johnson2010iterative}, are known to achieve excellent performance in practice in the classical
setting. Thus, the use of LDPC codes as the classical code in our construction,
decoded with iterative belief-propagation, BP-OSD
\cite{panteleev2021degenerate,roffe2020decoding}, or neural decoders, is a
topic of great interest.

\begin{acknowledgments}
I thank Kenneth Brown, Philip Johnson and Lorenza Viola for helpful comments. 
\end{acknowledgments}

\bibliographystyle{unsrt}

\clearpage

\onecolumngrid
\appendix

\section{\label{appback}Notation and background}
\subsection{Binary Linear Block Codes}\label{appclassical}
Error correcting codes protect messages against errors incurred during transmission
by adding redundant bits to the message. A code $\lcode \sim \left[ n, k, d \right]$  
is parameterized by the codeword length $n,$ message length $k,$ and  code distance $d.$
A binary, linear code $\lcode$ forms a $k$-dimensional subspace of $\mathbb{F}_2^n$
represented by $k$ linearly independent generator codewords
${\mathbf g}_1,{\mathbf g}_2, \dots , {\mathbf g}_k.$ Arranging the generators 
${\mathbf g}_i$ as rows of  a $k \times n$ binary generator matrix ${\mathbf G},$ a binary 
message ${\mathbf m}$ is mapped to a codeword by matrix multiplication ${\mathbf m} \gg.$
For binary codes, arithmetic is performed in $\mathbb{F}_2^n, $ that is, modulo 2. 
The generator matrix may be expressed in systematic form
$\gg = \left[ \,\pl \,\,|\, \, \id_{k}\,\right]$, where the $k \times (n-k)$ matrix $\pl$ defines a set 
of parity conditions that fix the redundant bits based on the message
to obtain  ${\mathbf m} \gg  = \left[ \, {\mathbf m}\pl \, |  \, {\mathbf m} \, \right]$. Note that for a 
generator matrix in systematic form, the transmitted codeword contains the redundant part and the message itself. 
The Hamming weight $\wt{{\mathbf v}}$ of a binary vector ${\mathbf v}$  is the number of its non-zero entries. 
A code's  distance $d$ is the defining parameter of the code since it specifies the minimum Hamming weight
of all the codewords and the minimum Hamming distance, defined as the number of places where two
binary vectors differ,  between any two codewords.  A distance $d$ code is capable of 
correcting all error patterns within the correction radius $t = \lfloor \frac{d-1}{2} \rfloor$,
but unable to correct all error patterns with $t+1$ errors.  

An alternative description of a linear code is given by the $(n-k) \times n$ parity-check matrix 
$\mathbf{H}$ defined as the orthogonal complement of ${\mathbf G}$ in $\mathbb{F}_2^n$, namely
$\mathbf{G}{\mathbf H}^\tp = {\mathbf 0},$ where ${\mathbf 0}$ is a $k \times (n-k)$ zero matrix.  
Viewed as a linear mapping $\mathbf{H}: {\mathbb F}_2^n \, \to \, {\mathbb F}_2^{n-k}$, the parity-check 
matrix maps a length $n$ vector to its $(n-k)$ length {\it syndrome}. 
 By definition, $\lcode$ is the kernel of  $\mathbf{H}$ 
and the quotient group ${\mathbb F}_2^n / \lcode$ is isomorphic to 
${\mathbb F}_2^{n-k}$. The cosets of ${\mathbb F}_2^n / \lcode$ can be arranged in a table called the {\it standard 
array} and used for syndrome decoding.  A  {\it coset leader}, 
defined as a minimum weight $n$-vector in the coset, is chosen as the coset representative for each coset.   A binary
vector ${\mathbf v}$ with $\wt{\mathbf v} \leq t$ is always a unique coset leader since the weight of any other
element of its coset ${\mathbf u} \in {\mathbf v} \oplus \lcode$, has weight $\wt{\mathbf u} > t$ since 
$\wt{\lcode} \geq d \geq 2t+1$.  By this property, $t$-error correcting codes can be used for data 
compression \cite{weiss1962compression}: a length $n$ binary vector is mapped to its 
$(n-k)$ length syndrome by $\mathbf{H}$. Provided the weight of the vector is 
 not greater than $t$, this mapping is invertible and the syndrome uniquely identifies the original vector.  

\subsection{Quantum Error Correcting Codes}\label{appquantum}
Let $\Pauli_\numQ = \left\langle \pm i I, X, Y, Z \right\rangle^{\otimes \numQ}$ 
denote the Pauli group on $\numQ$ qubits, where $XZ = -iY, XX=YY=ZZ=I,$ and $XZ=-ZX$.
The quotient group $\Pauli_\numQ/\left\{\pm i I\right\}$ is isomorphic to the binary vector space 
${\mathbb F}_2^{2\numQ}.$ 
Binary vectors $u, v \in \left\{0, 1\right\}^{\oplus \numQ } =  \mathbb{F}^\numQ_2$ 
define a general Pauli operator  by $\left[\, u \, | \, v \, \right] \definedby \Xu \Zv  = 
\bigotimes_{i=1}^{\numQ}X_i^{u_i} \bigotimes_{i=1}^{\numQ}Z_i^{v_i}$. 
Under the isomorphism, commutativity of  two Pauli operators is determined by the identity 
$$
\left[\, u \, | \, v \, \right]  \cdot \left[\, u' \, | \, v' \, \right] = (-1)^{u\cdot v' + u' \cdot v}
\left[\, u' \, | \, v' \, \right] \cdot \left[\, u \, | \, v \, \right]
$$
where $u\cdot v' = \oplus_i u_i v_i'$ is performed modulo 2. Thus, two Pauli operators
commute if the {\it symplectic} inner product \cite{calderbank1997quantum} ${u\cdot v' + u' \cdot v}$ vanishes. 

Recall the properties of a distance $d_Q$ quantum stabilizer 
code $\qcode$  encoding $\qk$ logical qubits in $\numQ$ physical qubits capable of 
correcting $t = \lfloor (\qd-1)/2\rfloor$ errors.  Codewords are simultaneous $+1$ 
eigenstates of $\qcode$'s stabilizer group $\stabgroup$, an abelian subgroup of the
Pauli group such that $\stab |\psi\rangle  = |\psi\rangle  $ for any $|\psi\rangle \in \qcode$
and  $\stab \in \stabgroup.$  
In the stabilizer formalism, detectable errors anti-commute with $\stabgroup$  and signal an 
error by observing a sign change in the stabilizer measurement outcomes.
The CSS  family of codes \cite{calderbank1996good,steane1996multiple} 
have stabilizer groups that are generated by distinct sets of either $X$ or $Z$-type Pauli operators, 
so that $\stabgroup = \langle \stabgroup_X, \stabgroup_Z\rangle$ where 
$\stabgroup_X = \{ \Xbeta: \beta \in S_X\} \, ,  
\stabgroup_Z = \{ \Zalpha: \alpha \in S_Z\}$ and $S_X, S_Z$ are
subspaces of $\mathbb{F}^\numQ_2.$ As linear subspaces, $S_X$ 
and $S_Z$ can be represented as matrices denoted, $\hqx$ and 
$\hqz$ respectively, with the binary representation of the $X$ and $Z$-type 
stabilizers arranged as rows.
For CSS codes, the symplectic inner product reduces to the modulo 2 inner product between
 $X$ and $Z$-type Pauli operators.
Error syndromes from $X$- and $Z$-type errors, $\Xu, \Zv,$ 
are obtained by the commutation identities 
$\Zalpha \Xu |\psi \rangle = (-1)^{\alpha \cdot u} \Xu \Zalpha |\psi\rangle$
and $\Xbeta \Zv   |\psi \rangle = (-1)^{\beta \cdot v} \Zv \Xbeta|\psi\rangle$. 
Bit and phase-flip 
recovery operations are performed separately in CSS codes and its parity-check 
matrix has the block structure
\be
\hq = 
\left[
\begin{array}{c c}
\hq^Z & \bf{0} \\ 
{\mathbf 0} & \hq^X
\end{array}
\right]
\ee
Under the isomorphism, error syndromes $\syndQ_X$ and $\syndQ_Z$ corresponding to  
Pauli errors $\Xu$ and $\Zv$, respectively,  can be formally obtained 
by modulo 2 matrix-vector  multiplication 
\be
\hqz u = \syndQ_X \quad \hqx v = \syndQ_Z
\ee

\subsubsection{$\steanecode$ Steane code}\label{appsteane} 
The $\steanecode$ Steane code \cite{steane1996multiple}
protects against a single bit-flip or phase-flip on $n=7$ physical qubits
encoding $k=1$ logical qubit.  The Steane code is a CSS code
constructed from the classical $\left[7, 4, 3 \right] $ Hamming 
code and its dual.  
\begin{table}[t]
\begin{center}
\begin{tabular}{cc}
\hline
   $s_X$& $s_Z$ \\
\hline
$X_{1}X_{4}X_{6}X_{7}$&$Z_{1}Z_{4}Z_{6}Z_{7}$\\ 
$X_{2}X_{4}X_{5}X_{7}$&$Z_{2}Z_{4}Z_{5}Z_{7}$\\ 
$X_{3}X_{4}X_{5}X_{6}$&$Z_{3}Z_{4}Z_{5}Z_{6}$\\
\hline
\end{tabular}
\end{center}
\caption{Generators of the stabilizer group of the Steane code}
\label{gen_stabilizer_steane}
\end{table}%
Pauli operators generating the stabilizer group are listed in Table \ref{gen_stabilizer_steane}.
The Steane code is a {\it dual-containing} code implying that $H_\qcode^Z = H_\qcode^X$.  
In the binary representation, with stabilizers arranged as rows, we have the quantum parity-check matrices
\be 
H_\qcode^Z = H_\qcode^X = 
\left[
\begin{array}{c c c c c c c}
1&0&0&1&0&1&1\\
0&1&0&1&1&0&1\\
0&0&1&1&1&1&0\\
\end{array}
\right]
\label{steane_parity}
\ee
The Steane code has distance 3, and the minimal weight generators of its
normalizer group, of weight 3, are listed in Table \ref{gen_normalizer_steane}. The weight-4 normalizers
are in the stabilizer, obtained by the product of all three generators of each error type in
Table \ref{gen_stabilizer_steane}.
\begin{table}[h]
\begin{center}
\begin{tabular}{cc}
\hline
   $\eta_X$& $\eta_Z$ \\
\hline
$X_{1}X_{2}X_{3}X_{4}$&$Z_{1}Z_{2}Z_{3}Z_{4}$\\ 
$X_{2}X_{3}X_{5}$&$Z_{2}Z_{3}Z_{5}$\\ 
$X_{1}X_{3}X_{6}$&$Z_{1}Z_{3}Z_{6}$\\ 
$X_{1}X_{2}X_{7}$&$Z_{1}Z_{2}Z_{7}$\\ 
\hline
\end{tabular}
\end{center}
\caption{Generators of the normalizer of the Steane code.}
\label{gen_normalizer_steane}
\end{table}%

The Steane code is non-degenerate, meaning that correctable error patterns have a unique
error syndrome.  A lookup table pairing error syndromes to error patterns is shown in
Table \ref{steane_lookup} in the case of bit-flips.
\begin{table}[h]
\begin{center}
\begin{tabular}{c c }
\hline
 100 & $X_{1}$  \\
 010 & $X_{2}$ \\ 
 001 & $X_{3}$ \\ 
 111 & $X_{4}$ \\ 
 011 & $X_{5}$ \\ 
 101 & $X_{6}$ \\ 
 110 & $X_{7}$ \\ 
\hline 
\end{tabular} 
\caption{Bit-flip lookup table for the Steane code. Weight-1 bit-flips (second column)
are indexed by the error syndrome determined from $\hq^Z.$ Phase flip syndromes are identical
but obtained independently by $\hq^X$ posing no ambiguity in the associated lookup table.}
\label{steane_lookup}
\end{center}
\end{table}

\subsubsection{$\colorcode$ color code}\label{appcolor}

Color codes \cite{PhysRevLett.97.180501, PhysRevA.81.032301}
are a class of topological QEC codes that allow 
for transversal implementations of the Clifford group. 
Here we present details for a distance 5 
code capable of correcting error patterns of weight-2  
$(t_\qcode = 2)$ or less with parameters $\colorcode.$  Color codes are CSS codes 
and therefore have stabilizer groups that partition into sets of  Pauli-$X$ 
or $Z$ operators only.  Syndrome measurements of Pauli-$X(Z)$ 
are used to detect $Z(X)$-type errors. 
\begin{figure}[t]
\begin{center}
\includegraphics[width=0.4\textwidth]{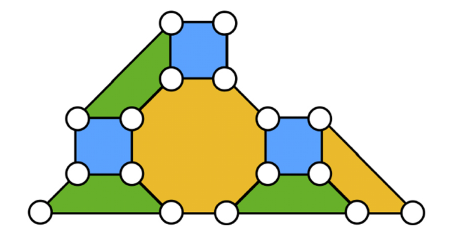}
\caption{ Qubit layout for the $\left[\left[ 17,1,5 \right]\right]$ color code
(reproduced from \cite{chamberland2017error}).
 Qubits are numbered left to right and top to bottom, e.g. data
 qubits 13--17 are in the bottom row. }
\label{planar_layout}
\end{center}
\end{figure}
Figure \ref{planar_layout} depicts the $\colorcode$ logical qubit. Each face defines a
stabilizer of $X$- and $Z$-type. In the binary representation, with  stabilizers arranged
 as rows, we have the quantum parity-check matrix
\be 
H_\qcode = 
\left[
\begin{array}{c c c c c c c c c c c c c c c c c}
1&1&1&1&0&0&0&0&0&0&0&0&0&0&0&0&0\\
1&0&1&0&1&1&0&0&0&0&0&0&0&0&0&0&0\\
0&0&0&0&1&1&0&0&1&1&0&0&0&0&0&0&0\\
0&0&0&0&0&0&1&1&0&0&1&1&0&0&0&0&0\\
0&0&0&0&0&0&0&0&1&1&0&0&1&1&0&0&0\\
0&0&0&0&0&0&0&0&0&0&1&1&0&0&1&1&0\\
0&0&0&0&0&0&0&1&0&0&0&1&0&0&0&1&1\\
0&0&1&1&0&1&1&0&0&1&1&0&0&1&1&0&0\\
\end{array}
\right]
\label{r_Q_planar}
\ee

As described in \cite{chamberland2017error}, encoding of CSS codewords may 
be performed with a row-reduced parity-check matrix without permuting the columns
(ordering of data qubits) as may be required to put the parity-check matrix in systematic form. 
A row-reduced parity-check matrix for  $H_\qcode$ is given by
\be
H_\qcode^r = 
\left[
\begin{array}{c c c c c c c c c c c c c c c c c}
1&0&0&1&0&1&0&0&0&1&0&0&1&0&1&1&1\\
0&1&0&1&0&0&0&0&0&0&0&0&1&1&0&0&0\\
0&0&1&1&0&1&0&0&0&1&0&0&0&1&1&1&1\\
0&0&0&0&1&1&0&0&0&0&0&0&1&1&0&0&0\\
0&0&0&0&0&0&1&0&0&0&0&1&0&0&1&0&1\\
0&0&0&0&0&0&0&1&0&0&0&1&0&0&0&1&1\\
0&0&0&0&0&0&0&0&1&1&0&0&1&1&0&0&0\\
0&0&0&0&0&0&0&0&0&0&1&1&0&0&1&1&0\\
\end{array}
\right]
\label{r_Q_rr}
\ee

Generators for the stabilizer and the normalizer  of the $\colorcode$ color code  
are shown in Table \ref{gen_17} and Table \ref{gen_norm_17}, respectively.
Since the minimum weight of the non-stabilizer elements of the normalizer is the distance of the code, from 
Table \ref{gen_norm_17}, we see that $\qd=5.$ 
\begin{table}[h]
\begin{center}
\begin{tabular}{cccc}
\hline
   $s_X$& $s_Z$ &$r_X$&$r_Z$\\
\hline
$X_{1}X_{2}X_{3}X_{4}$&$Z_{1}Z_{2}Z_{3}Z_{4}$&$X_{1}X_{4}X_{6}X_{10}X_{13}X_{15}X_{16}X_{17}$&$Z_{1}Z_{4}Z_{6}Z_{10}Z_{13}Z_{15}Z_{16}Z_{17}$\\ 
$X_{1}X_{3}X_{5}X_{6}$&$Z_{1}Z_{3}Z_{5}Z_{6}$&$X_{2}X_{4}X_{13}X_{14}$&$Z_{2}Z_{4}Z_{13}Z_{14}$\\ 
$X_{5}X_{6}X_{9}X_{10}$&$Z_{5}Z_{6}Z_{9}Z_{10}$&$X_{3}X_{4}X_{6}X_{10}X_{14}X_{15}X_{16}X_{17}$&$Z_{3}Z_{4}Z_{6}Z_{10}Z_{14}Z_{15}Z_{16}Z_{17}$\\ 
$X_{7}X_{8}X_{11}X_{12}$&$Z_{7}Z_{8}Z_{11}Z_{12}$&$X_{5}X_{6}X_{13}X_{14}$&$Z_{5}Z_{6}Z_{13}Z_{14}$\\ 
$X_{9}X_{10}X_{13}X_{14}$&$Z_{9}Z_{10}Z_{13}Z_{14}$&$X_{7}X_{12}X_{15}X_{17}$&$Z_{7}Z_{12}Z_{15}Z_{17}$\\ 
$X_{11}X_{12}X_{15}X_{16}$&$Z_{11}Z_{12}Z_{15}Z_{16}$&$X_{8}X_{12}X_{16}X_{17}$&$Z_{8}Z_{12}Z_{16}Z_{17}$\\ 
$X_{8}X_{12}X_{16}X_{17}$&$Z_{8}Z_{12}Z_{16}Z_{17}$&$X_{9}X_{10}X_{13}X_{14}$&$Z_{9}Z_{10}Z_{13}Z_{14}$\\ 
$X_{3}X_{4}X_{6}X_{7}X_{10}X_{11}X_{14}X_{15}$&$Z_{3}Z_{4}Z_{6}Z_{7}Z_{10}Z_{11}Z_{14}Z_{15}$&$X_{11}X_{12}X_{15}X_{16}$&$Z_{11}Z_{12}Z_{15}Z_{16}$\\ 
\hline
\end{tabular}
\end{center}
\caption{Generators of the stabilizer of the  $\colorcode$ color code in the planar basis $g$ (see FIG. \ref{planar_layout} 
and (\ref{r_Q_planar}) ) and the row-reduced basis $h$ (\ref{r_Q_rr})}
\label{gen_17}
\end{table}%
If a stabilizer has weight less than the code distance, the code is degenerate. From Table \ref{gen_17}, we
conclude that the $\colorcode$ code is degenerate since, for example, $\wt{X_{1}X_{2}X_{3}X_{4}} = 4 < \qd$.

Degenerate codes have multiple error patterns mapping to the 
same syndrome measurement which all lie in the same coset under the stabilizer group action.
A many-to-one mapping of error patterns to syndromes may introduce ambiguity in designing a decoder, nonetheless 
a lookup table for the $\colorcode$ may be tabulated by evaluating the syndromes for all low-weight $(\leq 2)$
error patterns.  There are $ 17 + {\numQ \choose 2} = 153 $ error patterns of weight less than or equal to $t_\qcode = 2.$
An enumeration of bit-flip cosets is shown in Table \ref{color_cosets}. Note that due to degeneracy, there are only 115
distinct syndromes corresponding to all weight-1 and 2 error patterns.
\begin{table}[h]
\begin{center}
\begin{tabular}{cc}
\hline
   $\eta_X$& $\eta_Z$ \\
\hline
$X_{1}X_{2}X_{3}X_{4}$&$Z_{1}Z_{2}Z_{3}Z_{4}$\\ 
$X_{1}X_{3}X_{5}X_{6}$&$Z_{1}Z_{3}Z_{5}Z_{6}$\\ 
$X_{1}X_{3}X_{9}X_{10}$&$Z_{1}Z_{3}Z_{9}Z_{10}$\\ 
$X_{7}X_{8}X_{11}X_{12}$&$Z_{7}Z_{8}Z_{11}Z_{12}$\\ 
$X_{1}X_{2}X_{5}X_{9}X_{13}$&$Z_{1}Z_{2}Z_{5}Z_{9}Z_{13}$\\ 
$X_{2}X_{3}X_{5}X_{9}X_{14}$&$Z_{2}Z_{3}Z_{5}Z_{9}Z_{14}$\\ 
$X_{1}X_{3}X_{7}X_{11}X_{15}$&$Z_{1}Z_{3}Z_{7}Z_{11}Z_{15}$\\ 
$X_{1}X_{3}X_{8}X_{11}X_{16}$&$Z_{1}Z_{3}Z_{8}Z_{11}Z_{16}$\\ 
$X_{1}X_{3}X_{7}X_{8}X_{17}$&$Z_{1}Z_{3}Z_{7}Z_{8}Z_{17}$\\ 
\hline
\end{tabular}
\end{center}
\caption{Generators of the normalizer of the $\colorcode$ color code. All weight-4 generators are elements of the stabilizer. }
\label{gen_norm_17}
\end{table}%

The coset table also serves as a lookup table decoder for the $\colorcode$ since any element 
of coset can correct any other element in the coset. Take, for example, the coset of errors
\be
E = \left\{ X_{1}X_{3}, X_{2}X_{4}, X_{5}X_{6}, X_{9}X_{10}, X_{13}X_{14} \right\}
\ee
which all trigger the syndrome measurement $``10100000".$ The product of any two elements of $E$
lies in the stabilizer and can therefore be corrected by a single Pauli-$X$ operator. For
example, the product 
\be
X_{1}X_{3}X_{9}X_{10}
\ee
can be written as a product of generators $r_1 r_3 r_7:$
\be
r_1 r_3 = X_{1}X_{3}X_{13}X_{14}
\ee
\bea
r_1 r_3 r_7 &=& X_{1}X_{3}X_{13}X_{14} X_{9}X_{10}X_{13}X_{14} \\ 
&=& X_{1}X_{3} X_{9}X_{10}
\eea
The correction $X_{1}X_{3}$ therefore corrects an error $X_{1}X_{3}$ but also the error $X_{9}X_{10}.$
Similar decomposition of products from $E$ may be computed and it can be shown that any element of
$E$ may be chosen as the corrective action to any other element of $E.$  
\begin{table}[t]
\begin{center}
\begin{tabular}{ccccccc}
\hline
 10000000 & $X_{1}$   & 01000000 & $X_{2}$ & 00100000 & $X_{3}$ &\\ 
 11100000 & $X_{4}$    & 00010000 & $X_{5}$ & 10110000 & $X_{6}$ & \\ 
 00001000 & $X_{7}$    & 00000100 & $X_{8}$ & 00000010 & $X_{9}$ & \\ 
 10100010 & $X_{10}$  & 00000001 & $X_{11}$ & 00001101 & $X_{12}$ &\\ 
 11010010 & $X_{13}$  & 01110010 & $X_{14}$ & 10101001 & $X_{15}$ &\\ 
 10100101 & $X_{16}$  &10101100 & $X_{17}$  & &  & \\
  11000000 & $X_{1}X_{2}$,$X_{3}X_{4}$ & 10100000 & $X_{1}X_{3}$,$X_{2}X_{4}$,$X_{5}X_{6}$,$X_{9}X_{10}$,$X_{13}X_{14}$ 
 & 01100000 & $X_{1}X_{4}$,$X_{2}X_{3}$ &\\ 
 00110000 & $X_{1}X_{6}$,$X_{3}X_{5}$ & 10001000 & $X_{1}X_{7}$ & 10000100 & $X_{1}X_{8}$ &\\ 
 10000010 & $X_{1}X_{9}$,$X_{3}X_{10}$ & 00100010 & $X_{1}X_{10}$,$X_{3}X_{9}$ & 10000001 & $X_{1}X_{11}$ &\\ 
 10001101 & $X_{1}X_{12}$ & 01010010 & $X_{1}X_{13}$,$X_{3}X_{14}$ & 11110010 & $X_{1}X_{14}$,$X_{3}X_{13}$ &\\ 
 00101001 & $X_{1}X_{15}$ & 00100101 & $X_{1}X_{16}$ & 00101100 & $X_{1}X_{17}$ &\\ 
 01010000 & $X_{2}X_{5}$,$X_{4}X_{6}$ & 11110000 & $X_{2}X_{6}$,$X_{4}X_{5}$ & 01001000 & $X_{2}X_{7}$ &\\ 
 01000100 & $X_{2}X_{8}$ & 01000010 & $X_{2}X_{9}$,$X_{4}X_{10}$ & 11100010 & $X_{2}X_{10}$,$X_{4}X_{9}$ &\\ 
 01000001 & $X_{2}X_{11}$ & 01001101 & $X_{2}X_{12}$ & 10010010 & $X_{2}X_{13}$,$X_{4}X_{14}$ &\\ 
 00110010 & $X_{2}X_{14}$,$X_{4}X_{13}$ & 11101001 & $X_{2}X_{15}$ & 11100101 & $X_{2}X_{16}$ &\\ 
 11101100 & $X_{2}X_{17}$ & 00101000 & $X_{3}X_{7}$ & 00100100 & $X_{3}X_{8}$ &\\ 
 00100001 & $X_{3}X_{11}$ & 00101101 & $X_{3}X_{12}$ & 10001001 & $X_{3}X_{15}$ &\\ 
 10000101 & $X_{3}X_{16}$ & 10001100 & $X_{3}X_{17}$ & 11101000 & $X_{4}X_{7}$ &\\ 
 11100100 & $X_{4}X_{8}$ & 11100001 & $X_{4}X_{11}$ & 11101101 & $X_{4}X_{12}$ &\\ 
 01001001 & $X_{4}X_{15}$ & 01000101 & $X_{4}X_{16}$ & 01001100 & $X_{4}X_{17}$ &\\ 
 00011000 & $X_{5}X_{7}$ & 00010100 & $X_{5}X_{8}$ & 00010010 & $X_{5}X_{9}$,$X_{6}X_{10}$ &\\ 
 10110010 & $X_{5}X_{10}$,$X_{6}X_{9}$ & 00010001 & $X_{5}X_{11}$ & 00011101 & $X_{5}X_{12}$ &\\ 
 11000010 & $X_{5}X_{13}$,$X_{6}X_{14}$ & 01100010 & $X_{5}X_{14}$,$X_{6}X_{13}$ & 10111001 & $X_{5}X_{15}$ &\\ 
 10110101 & $X_{5}X_{16}$ & 10111100 & $X_{5}X_{17}$ & 10111000 & $X_{6}X_{7}$ &\\ 
 10110100 & $X_{6}X_{8}$ & 10110001 & $X_{6}X_{11}$ & 10111101 & $X_{6}X_{12}$ &\\ 
 00011001 & $X_{6}X_{15}$ & 00010101 & $X_{6}X_{16}$ & 00011100 & $X_{6}X_{17}$ &\\ 
 00001100 & $X_{7}X_{8}$,$X_{11}X_{12}$,$X_{15}X_{16}$ & 00001010 & $X_{7}X_{9}$ & 10101010 & $X_{7}X_{10}$ &\\ 
 00001001 & $X_{7}X_{11}$,$X_{8}X_{12}$,$X_{16}X_{17}$ & 00000101 & $X_{7}X_{12}$,$X_{8}X_{11}$,$X_{15}X_{17}$ & 11011010 & $X_{7}X_{13}$ &\\ 
 01111010 & $X_{7}X_{14}$ & 10100001 & $X_{7}X_{15}$,$X_{8}X_{16}$,$X_{12}X_{17}$ & 10101101 & $X_{7}X_{16}$,$X_{8}X_{15}$,$X_{11}X_{17}$ &\\ 
 10100100 & $X_{7}X_{17}$,$X_{11}X_{16}$,$X_{12}X_{15}$ & 00000110 & $X_{8}X_{9}$ & 10100110 & $X_{8}X_{10}$ &\\ 
 11010110 & $X_{8}X_{13}$ & 01110110 & $X_{8}X_{14}$ & 10101000 & $X_{8}X_{17}$,$X_{11}X_{15}$,$X_{12}X_{16}$ &\\ 
 00000011 & $X_{9}X_{11}$ & 00001111 & $X_{9}X_{12}$ & 11010000 & $X_{9}X_{13}$,$X_{10}X_{14}$ &\\ 
 01110000 & $X_{9}X_{14}$,$X_{10}X_{13}$ & 10101011 & $X_{9}X_{15}$ & 10100111 & $X_{9}X_{16}$ &\\ 
 10101110 & $X_{9}X_{17}$ & 10100011 & $X_{10}X_{11}$ & 10101111 & $X_{10}X_{12}$ &\\ 
 00001011 & $X_{10}X_{15}$ & 00000111 & $X_{10}X_{16}$ & 00001110 & $X_{10}X_{17}$ &\\ 
 11010011 & $X_{11}X_{13}$ & 01110011 & $X_{11}X_{14}$ & 11011111 & $X_{12}X_{13}$ &\\ 
 01111111 & $X_{12}X_{14}$ & 01111011 & $X_{13}X_{15}$ & 01110111 & $X_{13}X_{16}$ &\\ 
 01111110 & $X_{13}X_{17}$ & 11011011 & $X_{14}X_{15}$ & 11010111 & $X_{14}X_{16}$ &\\ 
 11011110 & $X_{14}X_{17}$ & 10010000 & $X_{1}X_{5}$,$X_{3}X_{6}$ & && \\
\hline
\end{tabular} 
\end{center}
\caption{Lookup table for weight-1 and 2 bit-flips for the $\colorcode$ color code. }
\label{color_cosets}
\end{table}

\clearpage 

\section{Bit-flip  cosets and lookup table}\label{appbit}
\begin{table}[h]
\ZZIsorted
\caption{Cosets of $\Pauli_{12}$ under the action of $\pl^\tp \otimes \left[\, 1 \, 1 \, 0 \,\right]$}
\label{syndtablesZZI}
\end{table}

\begin{table}[h]
\ZIZsorted
\caption{Cosets of $\Pauli_{12}$ under the action of  $\pl^\tp \otimes \left[\, 1 \, 0 \,1\,\right]$}
\label{syndtablesZIZ}
\end{table}%

\begin{table}[h!]
\syndthree
\caption{Look-up table decoder for the three-qubit bit-flip code encoded 
with the parity-check matrix $\pl^\tp$ of the classical $\left[7,\,4,\,3 \right]$ 
Hamming code formed by joining Tables \ref{syndtablesZZI} and 
\ref{syndtablesZIZ}.  A unique syndrome exists for any single bit-flip occurring 
in the data qubits. Highlighted are the syndromes corresponding to 
bit-flips $X_2$ (red) and  $X_7$ (blue) and the weight-2 error $X_1X_{10}$ (green)
as depicted in the stabilizer circuits in Figures \ref{ZZI_circuit} and \ref{ZIZ_circuit}. }
\label{example_lookup}
\end{table}%

\end{document}